\documentclass[pdflatex,sn-mathphys-num]{sn-jnl}

\usepackage{graphicx}%
\usepackage{multirow}%
\usepackage{amsmath,amssymb,amsfonts}%
\usepackage{amsthm}%
\usepackage{mathrsfs}%
\usepackage[title]{appendix}%
\usepackage{xcolor}%
\usepackage{textcomp}%
\usepackage{manyfoot}%
\usepackage{booktabs}%
\usepackage{algorithm}%
\usepackage{algorithmicx}%
\usepackage{algpseudocode}%
\usepackage{listings}%
\usepackage{tikz}
\usepackage{tabularx}   
\usepackage{ragged2e}   
\usepackage{placeins}
\usetikzlibrary{arrows.meta,positioning,calc}

\usepackage{bm}
\usepackage{gensymb}
\usepackage{siunitx}

\usepackage{url}

\begin{document}

\title[Plasma wakefield: from accelerators to black holes]{Plasma wakefield: from accelerators to black holes}

\author*[1,2]{\fnm{Pisin} \sur{Chen}}\email{pisinchen@phys.ntu.edu.tw}

\author[1,2]{\fnm{Yung-Kun} \sur{Liu}}\email{r06222017@ntu.edu.tw}

\affil*[1]{\orgdiv{Department of Physics}, \orgname{National Taiwan University}, \orgaddress{\city{Taipei} \postcode{10617}, \country{Taiwan, ROC}}}

\affil[2]{\orgdiv{Leung Center for Cosmology and Particle Astrophysics}, \orgname{National Taiwan University}, \orgaddress{\city{Taipei} \postcode{10617}, \country{Taiwan, ROC}}}

\abstract{Commemorating the 2024 S. Chandrasekhar Prize, this review provides a retrospective on the genesis and evolution of plasma wakefield acceleration. It traces the journey from prehistory and the invention of the Plasma Wakefield Accelerator (PWFA), the establishment of its theoretical cornerstones, to its profound reverberations across fundamental physics, including astrophysics and analog gravity. The narrative emphasizes conceptual evolution, key theoretical breakthroughs, and future outlook, culminating in a vision for hybrid schemes and next-generation colliders. In addition to application to particle accelerators and high energy collider physics, it is found that plasma wakefield, with its ultra-intense acceleration, can also be applied to investigate gravity effects in the laboratory based on Einstein's equivalence principle. A specific example is accelerating flying relativistic plasma mirrors to investigate the celebrated black hole Hawking evaporation and the associated information loss paradox. We describe an ongoing experiment, AnaBHEL (Analog Black Hole Evaporation via Lasers), which aims at shedding some lights on the black hole information loss paradox.}


\maketitle

\footnote{This review article is based on the lecture delivered at the 2024 Chandrasekhar Prize ceremony in Malacca, Malaysia.}

\newpage 
\tableofcontents

\section{Introduction: The Quest for a New Paradigm in Particle Acceleration}

\subsection{Limitation of Conventional Accelerators}

For over a century, our understanding of the world has been deeply connected to the development of particle accelerators. From Rutherford's first artificial nuclear transmutation, the golden era of particle physics that discovers diverse fundamental particles culminated with the discovery of the Higgs boson at the Large Hadron Collider (LHC) that adds an important brick on the standard model of particle physics. Each leap in our understanding of the subatomic world has been enabled by a corresponding leap in accelerated particle energy. This progress has been largely driven by radio-frequency (RF) technology, where charged particles are accelerated by oscillating electromagnetic fields confined within metallic resonant cavities. However, this bedrock of high-energy physics is facing a fundamental limit.

The acceleration gradient in conventional RF cavities is constrained by the ``electrical breakdown" of the metallic materials. When the surface electric field on the cavity walls exceeds a material-dependent threshold, which is typically on the order of $\sim 100 \text{MV/m}$, unwanted field emission and plasma discharge occur, quenching the accelerating field and potentially damaging the structure \cite{Grudiev2009}. Such a "breakdown limit" poses a severe constraint on the energy gain per unit length. Consequently, pushing the energy frontier to the TeV scale and beyond necessitates the construction of colliders tens to hundreds of kilometers in length, such as the construction of 91-km Future Circular Collider (FCC) in Europe, leading to prohibitive costs and immense infrastructural challenges. Therefore, as earlier as the 1950s, people have already been attempting a paradigm shift: the development of alternative acceleration schemes capable of sustaining gradients orders of magnitude higher than what conventional structures can offer. The most promising path lies in harnessing the collective fields of a plasma, a fully ionized medium that, being already ``broken down", is immune to the material damage that destroys metallic accelerating structures.

\subsection{Prehistory of Plasma Acceleration}
Before the modern concept of plasma wakefield acceleration took shape, the notion of using the collective fields of a charge ensemble to accelerate particles had been investigated for decades. We refer to this era as the {\it Prehistory}, a period of intellectual gestation that laid the conceptual groundwork for the revolution to come. The term prehistory is chosen to imply the existence of a watershed moment, a {\it Dawn of History} that would later crystallize these nascent ideas into a cohesive and viable framework.

The concept of collective acceleration can be traced back to the early 1950s. Harvie, in an unpublished memo, first envisioned using the Coulomb field of an electron bunch to pull and accelerate protons \cite{Harvie1951}. Concurrently, Raudorf proposed an ``electronic ram" mechanism, where sudden stoppage of an electron beam would create a strong potential gradient for ion acceleration \cite{Raudorf1951}. These early thoughts, while rudimentary, contained the essential seed: replacing the externally applied fields of a cavity with the intrinsic, and potentially much stronger, fields generated by a collective of charged particles.

During this period, two main intellectual streams began to emerge. The first focused on using structured electron ensembles, such as rings or beams, to trap and accelerate ions. A pioneering vision was articulated by Alfvén and Wernholm in 1952, who proposed a ``collective-field accelerator", where a transversely focused electron beam would drag and accelerate heavy ions \cite{Alfven1952}. This line of thinking was significantly advanced by Budker, who, at the landmark 1956 CERN Symposium, proposed the use of a ``stabilized relativistic electron beam" in a ring configuration— the precursor to the Electron Ring Accelerator (ERA)—to confine and accelerate ions within its potential well \cite{Budker1956a,BudkerNaumov1956b}. These ideas spurred a vibrant research program and later found experimental validation, for instance, in the work of Graybill and Uglum, who observed ions being accelerated to multi-MeV energies by the deep potential well of a virtual cathode formed by an intense electron beam \cite{Graybill1968}. This stream powerfully demonstrated the feasibility of generating immense accelerating fields from collective electron structures.

The second parallel stream of thought, which would prove to be more directly ancestral to wakefield acceleration, focused on using plasma waves as the accelerating medium. The pivotal contribution that truly set the stage for this path was Veksler's ``coherent principle of acceleration," also presented at the 1956 CERN Symposium \cite{Veksler1956}. He articulated a profound insight: If a group of particles acts coherently as a single entity, the collective field it produces can be much more intense than the field from any individual particle. At the same symposium, Fainberg proposed using a plasma itself as an accelerating structure, suggesting that longitudinal waves within a plasma waveguide could serve this purpose \cite{Fainberg1956}. These two papers were remarkably prescient. Veksler provided the ``why", the principle of coherent collective action, while Fainberg suggested the ``what", the plasma medium.

In the subsequent two decades, these parallel ideas blossomed. On the beam-plasma front, seminal theoretical work by Hammer and Rostoker \cite{Hammer1970} and Lee and Sudan \cite{Lee1971} established a self-consistent understanding of how a high-current beam propagates in a plasma, demonstrating that the plasma return current could effectively neutralize the beam's self-magnetic field. This was crucial as it provided the theoretical basis for how a driver beam could even survive its journey through a plasma. On the laser front, a critical breakthrough came from Rosenbluth and Liu in 1972, who showed that the beating of two laser beams with a frequency difference close to the plasma frequency could resonantly drive a large-amplitude plasma wave \cite{Rosenbluth1972}, laying the foundation for what would become beat-wave acceleration.

In retrospect, the``Prehistory" was characterized by these two intertwining themes: the acceleration of ions via electron rings and virtual cathodes, and the conceptualization of using plasma waves for acceleration. Both lines of inquiry powerfully reinforced the central tenet that collective fields could shatter the limitations of conventional accelerators. The stage was set, waiting for the "historical" event that would synthesize these elements into the powerful and elegant concept of the plasma wakefield.

\begin{figure}[t]
\centering
\includegraphics[width=0.9\linewidth]{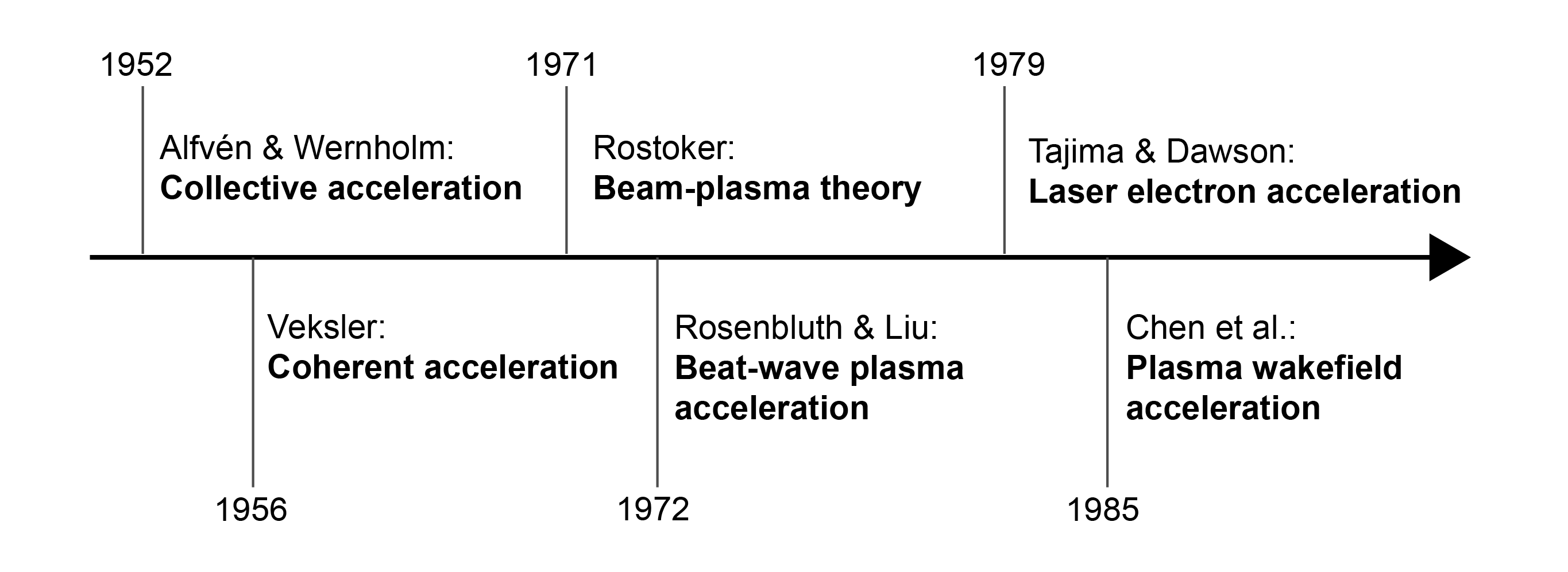}
\caption{From prehistory to the birth of wakefield acceleration (1952--1985).
We highlight key conceptual steps (collective/coherent acceleration and beam--plasma theory) that led to 
the 1979 \emph{laser-driven} milestones of PBWA and LWFA, and the 1985 \emph{beam-driven} breakthrough
that marks the birth of PWFA.}
\label{fig:timeline_core}
\end{figure}

\subsection{The Laser-Ignited Revolution}

The rich tapestry of ideas woven during the ``Prehistory" set a fertile ground for a breakthrough. That breakthrough arrived in 1979 with the seminal paper by Toshiki Tajima and John M. Dawson, "Laser Electron Accelerator" \cite{Tajima1979}. This work represents the "Dawn of History", a moment when the diffuse concepts of using plasma for acceleration were synthesized into a concrete, physically realizable, and astonishingly powerful scheme.

Tajima and Dawson's central insight was to recognize that an intense, short-pulse laser could be the ideal driver for exciting a large-amplitude plasma wave. They showed that the non-linear ponderomotive force, $\mathbf{F}_p = -(e^2/4m_e\omega_0^2)\nabla |\mathbf{E}_0|^2$, exerted by a laser pulse on plasma electrons acts like a snowplow. As the pulse propagates through the plasma at a velocity near the speed of light, $c$, it expels electrons from its path, both forward and sideways. These displaced electrons, feeling the restoring force from the background ions, rush back towards the axis behind the pulse, overshoot, and begin to oscillate at the electron plasma frequency, $\omega_p = (4\pi n_e e^2/m_e)^{1/2}$. This process establishes a coherent, relativistic plasma wave that trails the laser pulse.

The longitudinal electric field associated with this plasma wave can be immense. The maximum, or ``wave-breaking," field is given by $E_{\text{WB}} \approx m_e c \omega_p / e$, which scales with the square root of the plasma density $n_e$. For typical plasma densities used in experiments (e. g., $n_e \sim 10^{18}\,\text{cm}^{-3}$), this field can reach $\sim 100\,\text{GV/m}$, three orders of magnitude beyond the breakdown limit of conventional RF cavities. The Tajima-Dawson paper demonstrated, through theory and one-dimensional particle-in-cell (PIC) simulations\cite{Dawson1983,BirdsallLangdon1991}, that a second, properly phased bunch of electrons could "surf" on this plasma wake and be accelerated to ultra-high energies over very short distances. This concept, later referred to as the Laser Wakefield Accelerator (LWFA), ignited a worldwide research effort that continues to flourish to date.

\subsection{Consolidation of Revolution with Particle Beams}

The laser-driven scheme proposed by Tajima and Dawson was revolutionary. However, in the early 1980s, the laser technology required to achieve the necessary intensities (Chirped Pulse Amplification \cite{Strickland1985}) was yet to be born. In contrast, high-energy particle accelerators have already been producing intense, relativistic particle beams for decades. This context prompted a natural and pivotal question: Would a relativistic charged particle beam serve as the driver to excite large amplitude plasma waves, in a manner analogous to a laser pulse?

This question was answered in the affirmative through another seminal work, put forward by Chen et al. \cite{Chen1984,Chen1985}, which first showed, through a concise linear analysis and PIC simulations, that a relativistic electron beam can drive a Langmuir wave in the plasma with large amplitude, via its space-charge field. The plasma response can be expressed as a convolution of the bunch charge density with the plasma Green-function, yielding simple analytic formulas for the accelerating field. This work demonstrates that, although the underlying plasma dynamics are different, the plasma waves excited by particle beams can be as large as that by lasers. 

Upon learning about this concept from the authors \cite{Chen1984,Chen1985}, Ruth et al. \cite{Ruth1985} recognized its analogy with the ``wakefield accelerator" concept in accelerator physics, coined the term ``Plasma Wakefield Accelerator (PWFA)" for the concept proposed by Chen et al., and carved it in the language of wakefield theory in accelerator physics. They introduced the now-standard driver/witness beam configuration and investigated key beam dynamics issues in PWFA, including beam-loading and energy spread, transverse-focusing-induced betatron motion, beam emittance preservation, etc. One important conclusion in this investigation is that PWFA obeys the known Fundamental Theorem of Beam Loading \cite{Wilson1982} in accelerator beam physics, which dictates that the energy gain of the accelerating beam cannot be greater than twice the energy of the driving beam. Incidentally, since the term ``wakefield" was rapidly adopted, the advanced accelerator research community later referred to the Tajima-Dawson concept as ``Laser Wakefield Accelerator (LWFA)". This foundational work in the 1980s set the stage for decades of research that have culminated in the advanced, facility-scale programs operating today. For a comprehensive overview of the current state-of-the-art in beam-driven plasma wakefield acceleration, encompassing the latest theoretical understanding, experimental breakthroughs, and future facility roadmaps, we direct the reader to the recent extensive review by Lindstrøm et al.\cite{Lindstrøm2025}.

\section{The Physical Foundation of PWFA: Laying the Theoretical Cornerstone}

The conception of Plasma Wakefield Acceleration (PWFA) in 1985 introduced a paradigm shift in accelerator physics, replacing conventional metallic cavities with dynamic, self-generated waves in a plasma \cite{Chen1984,Chen1985,Ruth1985}. This section delineates the theoretical foundations of PWFA, commencing with the fundamental process of wakefield excitation and subsequently addressing key concepts and challenges, including transformer ratio, beam loading, and transverse beam dynamics. 

\subsection{The Wakefield Principle}

At its core, PWFA operates by transferring energy from a relativistic driving particle beam to plasma Langmuir waves, which in turn can accelerate a trailing witness beam. The coupling mechanism is the space-charge force of the driver, in contrast to the ponderomotive force that drives Laser Wakefield Accelerators (LWFA). The plasma response in PWFA is critically dependent on the intensity of the perturbation. This intensity is quantified by the ratio of the driver's charge density, $n_b$, to the ambient plasma density, $n_p$, a parameter denoted by $\phi_0$:
\begin{equation}
\phi_0 \equiv \frac{n_b}{n_p}.
	\label{eq:phi0}
\end{equation}
This parameter plays a role analogous to the normalized laser strength parameter, $a_0$, in LWFA. Whereas $a_0$ represents the normalized vector potential of the electromagnetic perturbation, $\phi_0$ signifies the normalized scalar potential. The thresholds of $a_0 \approx 1$ and $\phi_0 \approx 1$ delineate the transition from the linear regime ($a_0, \phi_0 \ll 1$) to the non-linear regime ($a_0, \phi_0 \gtrsim 1$). In this non-linear regime, the plasma electron motion becomes strongly relativistic, as the energy gained by electrons over a characteristic length scale (e.g., a laser cycle for LWFA) becomes comparable to their rest mass energy. Consequently, effects such as the relativistic mass increase and the $\mathbf{v} \times \mathbf{B}$ force component become dominant, leading to highly non-sinusoidal wake structures.

Although the underlying dynamics of PWFA and LWFA may be different, they both obey the so-called {\it Wakefield Principle}, that is,

\begin{center}
\textbf{The phase velocity of a wake in a medium is equal to} 

\textbf{the group velocity of the entity that excites it.}
\end{center}
This principle is general, applying not only to plasmas but also to other media, including conventional RF cavities, dielectric structures, and even the wake behind a boat. It holds true regardless of whether the perturbation is in the linear or non-linear regime, making it a cornerstone of wakefield physics.

A direct consequence of this principle is that the wake's spatial period, or wavelength $\lambda_w$, is determined by both the driver's velocity and the plasma's response. In the laboratory frame, the wake wavelength can be expressed as:
\begin{equation}
    \lambda_w = \frac{2\pi v_{\text{ph}}}{\omega_{\text{eff}}},
    \label{eq:lambda_general}
\end{equation}
where $\omega_{\text{eff}}$ is the effective oscillation frequency of the plasma electrons. This frequency is modified by relativistic effects. We can capture this by defining $\omega_{\text{eff}} = \omega_p / \sqrt{\Gamma}$, where $\omega_p \equiv (n_p e^2 / \epsilon_0 m_e)^{1/2}$ is the non-relativistic plasma frequency and $\Gamma$ is a relativistic factor:
\begin{equation}
\Gamma \equiv
\begin{cases}
1 & \text{(linear regime, cold plasma)}\\
\langle \gamma_e \rangle & \text{(nonlinear regime, relativistic electron motion)}
\end{cases}
\label{eq:gamma_factor}
\end{equation}
Here, $\langle \gamma_e \rangle$ is the average Lorentz factor of the oscillating plasma electrons. In the linear regime ($\Gamma=1$), the wake wavelength $\lambda_w \approx \lambda_p = 2\pi c/\omega_p$, with small corrections due to the driver's velocity being slightly less than $c$. In the nonlinear regime ($\Gamma = \langle\gamma_e\rangle > 1$), the relativistic mass increase of the plasma electrons leads to a lower effective oscillation frequency, causing a "stretching" of the wake wavelength, $\lambda_w^{\text{(NL)}} \approx \lambda_p \beta_{\text{ph}} \sqrt{\langle\gamma_e\rangle}$, where $\beta_{\text{ph}}=v_{\text{ph}}/c$. This lengthening of the wake period is a key signature of nonlinearity in both PWFA and LWFA. The magnitude of this stretching is directly related to the driver's intensity. For instance, for a linearly polarized laser driver in LWFA, the average electron Lorentz factor can be approximated as $\langle\gamma_e\rangle \approx \sqrt{1+a_0^2/2}$, leading to a wavelength of $\lambda_w^{\text{(LWFA,NL)}} \approx \lambda_p (1+a_0^2/2)^{1/4}$. Similarly, for PWFA, $\langle\gamma_e\rangle$ is determined by the amplitude of the normalized wake potential, $\hat{\psi} \equiv e\phi/m_e c^2$. This nonlinear lengthening of the wake period is a key experimental signature of the transition from the linear to the high-gradient nonlinear regime.

\begin{figure}[h!]
    \centering
    \includegraphics[width=0.9\linewidth]{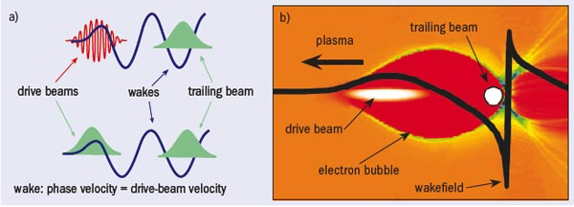}
    \caption{Comparison of laser-driven (LWFA) and beam-driven (PWFA) wake excitation. (a)~Schematic of wake generation: in LWFA (top), a high-intensity laser pulse drives the wake via the ponderomotive force; in PWFA (bottom), a relativistic particle bunch drives the wake via its space-charge field. In both cases, a trailing witness beam can be injected into the accelerating phase of the wake.     (b)~Particle-in-cell simulation snapshot of the nonlinear bubble regime in PWFA, showing the drive beam, trailing beam, electron-depleted cavity, and the surrounding wakefield. Adapted from \cite{joshi2007}.}
    \label{fig:LWFA_PWFA_schematic}
\end{figure}

Figure~\ref{fig:LWFA_PWFA_schematic} schematically contrasts the two approaches: in LWFA (panel a, top), an intense laser pulse excites a plasma wave whose phase velocity matches the laser group velocity; in PWFA (panel a, bottom), a high-charge particle bunch excites a wake at the beam velocity. Panel b illustrates the non-linear \emph{bubble} regime, in which the drive beam fully evacuates plasma electrons, forming a spherosoidal ion cavity that focuses and accelerates the trailing beam.

\subsubsection{The Linear Regime: An Intuitive Picture}
\label{sec:linear_regime}

The simplest and most analytically tractable description of wakefield excitation emerges in the linear regime, defined by the condition $n_b \ll n_p$. In this limit, the perturbation to the plasma is small, and the plasma electrons can be modeled as simple harmonic oscillators that respond linearly to the driver's electrostatic field.

As a relativistic driver bunch with charge density $\rho_b = -e n_b$ traverses the plasma, its space-charge field repels the plasma electrons radially outward. The ions, approximately 2000 times more massive for hydrogen plasma, are considered stationary on the timescale of the electron motion and form a fixed, neutralizing background. Once the driving force from a particular segment of the bunch has passed, the displaced plasma electrons are pulled back towards their equilibrium position by the electrostatic restoring force of the ion column. Their inertia causes them to overshoot the axis, initiating collective oscillations at the characteristic electron plasma frequency, $\omega_p = \sqrt{4\pi n_p e^2 / m_e}$. This collective oscillation constitutes the plasma wakefield.

The dynamics is described by the linearized cold–fluid equations for the plasma electrons.  
Under the quasistatic approximation, appropriate for a highly relativistic beam ($\gamma\gg1$) propagating along the $z$ axis, both the longitudinal wakefield $E_z$ and the transverse wakefield $W_\perp\equiv E_r-B_\theta$ can be derived from a single wake potential $\Psi\equiv\phi-A_z$,\cite{Katsouleas1986PRA,Katsouleas1987}:
\begin{equation}
	\bigl(\nabla_\perp^{2}-k_p^{2}\bigr)\,\Psi(\mathbf{x}_\perp,\zeta)
	\;=\;-\,4\pi\,\rho_b(\mathbf{x}_\perp,\zeta),
	\label{eq:wake_potential}
\end{equation}
where $\zeta=z-ct$ is the co-moving coordinate, $k_p=\omega_p/c$ is the plasma wavenumber, and $\phi$ and $A_z$ are the scalar and vector potentials, respectively.  
Equation~\eqref{eq:wake_potential} is a forced Helmholtz equation: each transverse `slice' of the driver excites a sinusoidal wake that trails behind it.  
For a transversely uniform (1-D) bunch, $\nabla_\perp^{2}\!\to\!\partial_\zeta^{2}$ and the equation reduces to the familiar simple harmonic form
\begin{equation}
	\bigl(\partial_\zeta^{2}+k_p^{2}\bigr)\,\Psi(\zeta)
	\;=\;-\,4\pi\,\rho_b(\zeta),
\end{equation}
So, the total wakefield is the linear superposition of the contributions from all beam slices, an idea captured by the Green function method introduced in \cite{Chen1985}.

A key feature of the linear wake is its fixed oscillation wavelength, the plasma wavelength $\lambda_p = 2\pi/k_p$. This sets the fundamental length scale of the accelerator: the driver and witness bunches must be significantly shorter than $\lambda_p$, and their separation must be precisely controlled to position the witness beam at a phase of maximum acceleration. While the longitudinal field provides acceleration, the co-existing transverse field is equally critical for beam confinement. As explored in a subsequent section, the structure and uniformity of this transverse wake are paramount in preserving the quality of the accelerated beam.

\begin{figure}[h!]
    \centering
    \includegraphics[width=0.7\linewidth]{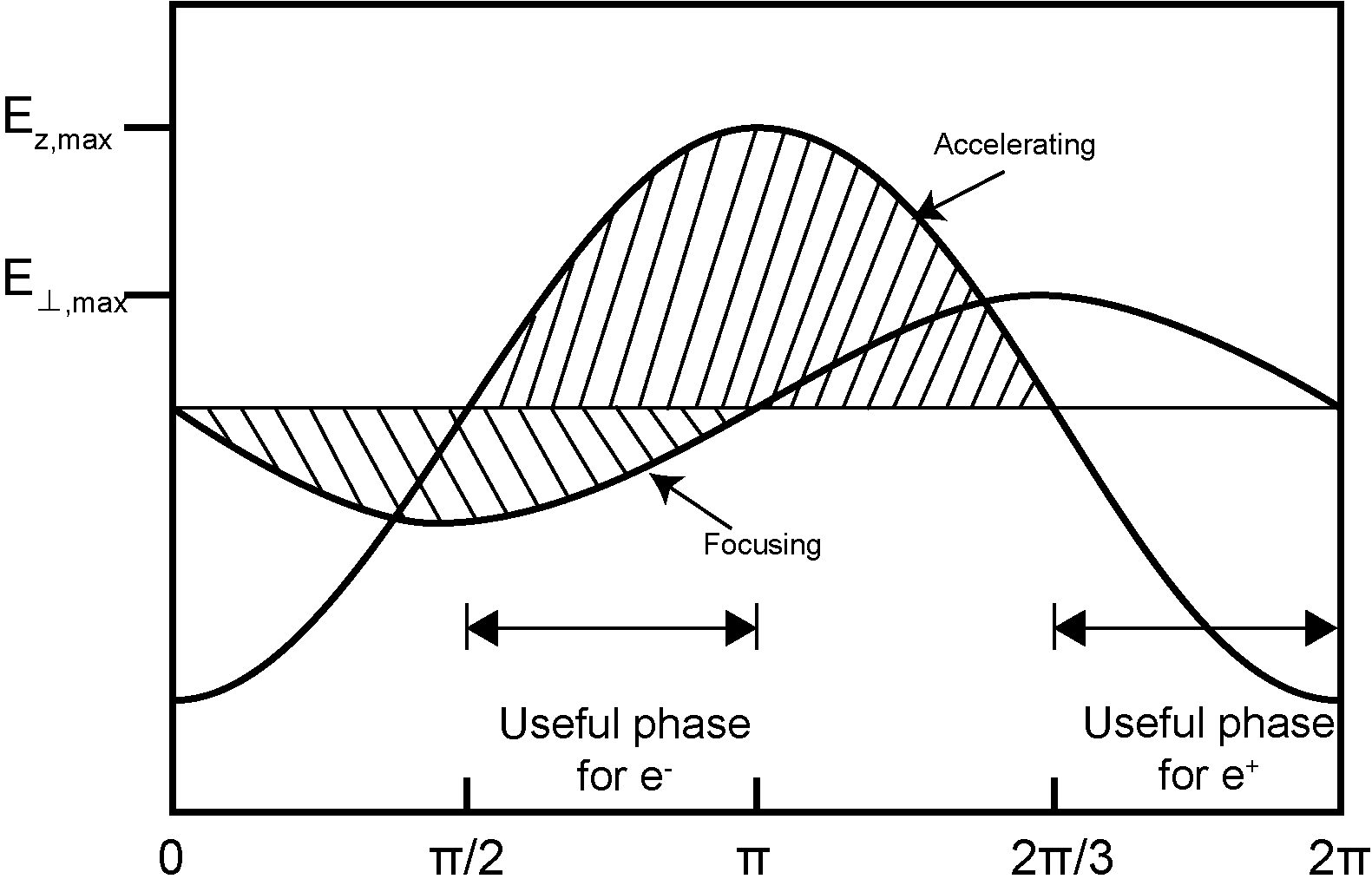} 
    \caption{Schematic of the longitudinal accelerating field ($E_z$) and transverse focusing field ($E_\perp$) in a linear plasma wakefield. The shaded regions indicate the phases where the force is accelerating or focusing for a trailing electron. For electrons (e$^-$), a "useful phase" of length $\pi/2$ exists where both acceleration and focusing occur. A similar stable region exists for positrons (e$^+$), but shifted in phase. Reproduced from~\cite{Ruth1986}.}
    \label{fig:linear_wake_phase}
\end{figure}

The interplay between the longitudinal and transverse fields determines the "useful phase" for particle acceleration. As illustrated in Fig.~\ref{fig:linear_wake_phase}, in a linear wake driven by an electron bunch, the longitudinal accelerating field ($E_z > 0$) and the transverse focusing field ($W_\perp < 0$) are $\pi/2$ out of phase. This phase relationship creates a quadrant of the plasma wavelength where particles can experience both acceleration and focusing simultaneously. It should be noted that this property holds for both electrons and positrons, albeit in different phase regions of the wake. This fundamental characteristic underpins the viability of plasma accelerators for both particle species in the linear regime, although as we will see, it presents significant challenges for positrons in the non-linear regimes.

\subsubsection{The Nonlinear Regime: Towards Extreme Gradients}

Although linear theory provides an invaluable analytical framework, pushing to the highest accelerating gradients inevitably requires operation in the nonlinear regime.  Nonlinearity dominates once the density of the driving beam approaches or exceeds the plasma density ($\phi_0=n_b/n_p \gtrsim 1$) and the beam is transversely narrow ($k_p\sigma_r\!\lesssim\!1$), conditions under which the driver’s space-charge force expels virtually all plasma electrons from its path.  The resulting electron-free ion column is the hallmark of the so-called \emph{blow-out} (or \emph{bubble}) regime~\cite{Rosenzweig1991,Lu2006}.

Early 1-D non-linear analyzes showed that as the wave amplitude increases, the oscillation frequency becomes amplitude dependent and the waveform steepens from a sinusoid to a sawtooth shape, ultimately 'breaking' in the cold plasma wavebreaking field $E_{\text{WB}} = m_e c \omega_p / e$~\cite{Akhiezer1956,Dawson1959,Coffey1971}. While this indicated the possibility of higher gradients, the crucial multi-dimensional physics of beam-driven wakes was not included.

Full 2-D/3-D theory and PIC simulations reveal that a dense, short electron beam carves out a nearly spherical ion cavity, or bubble, whose properties are highly favorable for acceleration:
\begin{enumerate}
	\item \textbf{Extreme accelerating field.}
	The electron sheath at the bubble boundary and the uniform ion column establish a strong longitudinal field $E_z$ at the back of the bubble that approaches, and in favorable cases modestly exceeds, $E_{\text{WB}}$.
	\item \textbf{Linear focusing.}
	Within the ion column the transverse restoring force is $F_r = 2\pi n_p e^2 r$ (cgs), perfectly linear in radius $r$, providing an ideal channel for preserving witness-beam emittance.
	\item \textbf{Field separation.}
	The strongest accelerating field is localized near the tail of the bubble, whereas the linear focusing force pervades the entire cavity, effectively decoupling focusing from acceleration.
\end{enumerate}

The transition from linear to nonlinear bubble regime represents a qualitative change in the wakefield structure, not merely a quantitative increase in acceleration gradient. The bubble effectively acts as a self-formed, ideally shaped accelerating cavity composed of plasma. Modeling this regime is analytically complex, relying heavily on phenomenological models benchmarked by large-scale Particle-In-Cell (PIC) simulations \cite{Lu2006}. The successful generation and utilization of this bubble regime in experiments at facilities like SLAC FACET \cite{Litos2014} constituted a landmark achievement, confirming that the extreme gradients promised by nonlinear theory are physically attainable. The pristine focusing properties of both the linear wake and the nonlinear bubble are so fundamental to beam quality preservation that they warrant a dedicated review.

\subsection{Theorem of Ultimate Transformer Ratio: Evading the Limit set by Fundamental Theorem of Beam Loading}

Generating a large-amplitude wakefield is a necessary but insufficient condition for a practical accelerator. The ultimate goal is to efficiently transfer energy from the driving beam to the accelerating beam. This raises a critical question, rooted in the energy conservation arguments long familiar in conventional RF accelerators \cite{Wilson1982}:  
What is the maximum energy an accelerating beam particle can gain relative to the energy lost by the driving beam particles? A figure-of-merit that addresses this question is the \textbf{transformer ratio},~$R$.

Formally, $R$ is defined as the ratio of the peak accelerating field directly behind the driver to the peak decelerating field within it,

\begin{equation}
	R \equiv \frac{|E_{z, \text{accel}}|}{|E_{z, \text{decel}}|},
	\label{eq:transformer_ratio}
\end{equation}

and a large value of $R$ implies that a modest-energy driver can boost a witness bunch to much higher energies in a \emph{single} stage. For a single, longitudinally \emph{symmetric} bunch driving a linear wake, the ``Fundamental Theorem of Beam Loading" (FTBL) enforces an upper bound  $R\le2$\ \cite{Ruth1985,Bane1985}. Originally derived for metallic structures, the FTBL theorem should be equally applicable to plasmas in the linear regime and posed a severe barrier to TeV-class acceleration within one plasma stage.

The key insight, articulated in 1985–1986, is that the $R\!\le\!2$ bound is not caused by fundamental physical laws but by the assumption of head-tail symmetry in the longitudinal density profile of the driving beam. Bane, Chen and Wilson first showed that a triangular, ramped current profile can exceed the limit set by FTBL,\cite{Bane1985}. In the same article, the authors went further to prove a new theorem, {\it Theorem of Ultimate Transformer Ratio}, in effect replacing FTBL. 

In principle, every point-particle would induce a wakefield.  The longitudinal wake potential of the entire beam is therefore describable as the convolution of the beam density distribution with the Green function deriving from the point-particle perturbation,
\begin{align}
 \Psi(\zeta)=\int_{-\infty}^{\infty} G(\zeta-\zeta')\,\rho_b(\zeta')\,\mathrm{d}\zeta',   
\end{align}
where $G(\zeta-\zeta')$ is the Green function of the plasma response. Because the decelerating field at $\zeta$ inside the driver samples only the convoluted wakefield from the head to the location, whereas the accelerating field behind the beam samples the integrated wakefield of the entire bunch, an asymmetric profile naturally yields $E_{z,\mathrm{accel}}\!>\!|E_{z,\mathrm{decel}}|$.

In their foundational 1985 analysis, Bane, Chen and Wilson went further: they recognized that the most efficient way to transform beam energy to plasma wakefield is to extract energy, or deceleration, uniformly. One important consequence is that the driving beam would preserve its shape, which in turn would render the longest possible distance of propagation, and therefore more efficient energy extraction. Another way to appreciate this effect is that the ultimate beam profile that renders uniform deceleration of all beam particles manages to suppress the micro-bunching instability. Incidentally, the same philosophy is applicable to laser-plasma interactions if suppression of micro-bunching instability is desired. 

Using the inverse Laplace transform technique, Bane, Chen and Wilson solved for the beam current profile that produces a \emph{constant decelerating field} throughout the bunch.  Starting from the definition of wake potential,
\begin{align}
V(t) &= -\int_{-\infty}^{\infty} I(t')\,\Psi(t-t')\,\mathrm{d}t',
\end{align}
they obtain the corresponding driving beam current as
\begin{align}
I(t) = \frac{1}{2\pi i}\int_{\gamma - i\infty}^{\gamma + i\infty}
   \frac{\mathcal{L}\{V(t)\}(s)}{\mathcal{L}\{W_z(t)\}(s)}\,e^{st}\,\mathrm{d}s ,
   \label{eq:laplace_inversion}
\end{align}
where $\mathcal{L}$ denotes the Laplace transform and $\gamma$ is some positive quantity to avoid integration along the imaginary axis of the complex plane. Applying this inversion to the case of a constant retarding potential across the bunch yields the explicit solution
\begin{align}
I(t) = -\frac{V_0}{2k
\alpha}\left[(\alpha^2+\omega^2)e^{-\alpha t} + \omega^2(\alpha t - 1)\right],
\quad 0 < t < T,
\label{eq:optimal_current}
\end{align}
which can be viewed as a superposition of an exponentially decaying component and a triangular ramp. In the asymptotic limit of $\alpha \to \infty$, the exponential reduces to a delta-function at the head of the bunch, and the triangular term dominates. This “constant-deceleration” solution yields 100\% efficiency and the ultimate transformer ratio for a single-mode structure or medium:
\begin{equation}
  R_{\infty} \;\xrightarrow{\omega\rightarrow\infty}\; \sqrt{1+(2\pi N)^2} .
\end{equation}
This proves the Theorem of Ultimate Transformer Ratio.

This concept of using an asymmetric driver to enhance the transformer ratio proved to be remarkably universal. Theoretical work soon extended the idea from particle beams to lasers in the Laser Wakefield Accelerator (LWFA). Although the driving mechanism is different (ponderomotive vs. space-charge force), the guiding physical principle of uniformly decelerate all particles in the driving beam (PWFA) and uniformly redshift all photons in the laser (LWFA) to maximize energy transfer efficiency is the same. A series of papers by Chen and Spitkovsky established the theoretical framework for this extension to the nonlinear regime, showing that by tailoring the longitudinal envelope of the driving laser pulse— for example, using a slow rise and a sharp fall-off— an enhanced transformer ratio could be achieved for LWFA \cite{Chen1998, ChenSpitkovsky1999AIP}. They further developed a general optimization formalism and proposed practical shaping methods to achieve a state of "uniform photon deceleration," the optical analog to the constant decelerating field in the optimal PWFA case \cite{SpitkovskyChen2001AIP, SpitkovskyChen2002PLA}. This work underscored the deep generality of the underlying physics.

The principle of driving beam shaping has since become a central theme in advanced accelerator concepts. While parallel work, such as that proposal by Voss and Weiland at DESY, explored using a train of discrete bunches to build up a wakefield \cite{Voss1982}, the single-shaped-driver concept offered an elegant, fully collinear solution. It has profoundly guided R\&D, spurring the development of sophisticated beam-shaping techniques and yielding experimentally measured transformer ratios well in excess of the limit of $R\leq 2$, with values of $R \approx 4.6$ at DESY, $R \approx 5$ in a dielectric structure at Argonne Wakefield Accelerator (AWA) facility, and up to $R \approx 7.8$ in a nonlinear PWFA at AWA \cite{Loisch2018, Gao2018PRL, Roussel2020}.

The picture becomes even more favorable when the driver is intense enough to \emph{blow out} all plasma electrons and form an ion cavity. In this highly nonlinear regime, the simple linear theory and the formal $R \le 2$ bound are no longer applicable. The accelerating field is supplied by the highly nonlinear bubble wake, and the optimal driver is no longer necessarily a very long, gentle ramp. Instead, a moderately-long, high-current, ramped bunch can provide the best energy-transfer efficiency. Lotov showed analytically and with quasistatic particle-in-cell simulations that a driver with a length of $L_b \sim 2\lambda_p$ can yield $R \approx 4$ with over 90\% efficiency \cite{Lotov2005}. The aforementioned experiments have since confirmed and surpassed these predictions, demonstrating that shaped drivers in the blowout regime can deliver both multi-GV/m gradients \emph{and} high transformer ratios. This firmly establishes high transformer ratio operation as a realistic path toward compact, high-energy accelerators.

\subsection{Beam Loading}

The breakthrough of achieving a high transformer ratio with shaped drivers addressed the efficiency of wakefield generation. The next logical consideration is the back-reaction of the witness beam to the wakefield, a process known in accelerator beam physics as "beam loading",  The witness current generates its own wake, which superimposes on the driver's wake, potentially altering the accelerating field.

Optimal beam loading involves injecting the right amount of charge with the correct longitudinal profile to achieve two goals: maintaining a nearly constant accelerating field $E_z$ along the bunch for a small final energy spread, and extracting as much energy as possible from the wake to maximize efficiency. If the witness beam charge is too high (over-loading), the net $E_z$ can flip sign, causing the tail particles of the witness bunch decelerated. Conversely, under-loading leaves a residual wake, typically resulting in a larger energy spread. 

The energy transfer efficiency can be defined as $\eta\equiv\Delta W_{\text{witness}}/\Delta W_{\text{driver}}$.
This process was first studied theoretically by Ruth et al. \cite{Ruth1985} and via 1D PIC simulations by Chen et al. \cite{Chen1986} and Katsouleas \cite{Katsouleas1986PRA} in 1986, which established the beam-loading vocabulary and pointed towards the higher efficiencies achievable in the nonlinear regime. Subsequently, Katsouleas et al. elucidated the fundamental trade-off between efficiency, energy spread, and accelerating gradient in the linear regime, proposing a triangular witness beam profile to optimize both efficiency and energy spread \cite{Katsouleas1987_beamLoading}. Early linear theory models suggested a trade-off between high transformer ratios and high efficiency. However, this "efficiency curse" was shown via simulations by Lotov in 2005 not to be a fundamental limit in the nonlinear blowout regime. He pointed out that an "efficient operating mode" using a ramped driver and a trapezoidal witness beam in the blowout regime could push $\eta$ beyond $50\%$ while maintaining a low energy spread \cite{Lotov2005}.

A complete and self-consistent nonlinear theory of beam loading was presented by Tzoufras et al. in 2008 \cite{Tzoufras2008}. In the bubble regime, the longitudinal field $E_z$ is determined by the bubble's size and shape. By tailoring the witness beam profile, which in turn modifies the bubble shape, the longitudinal field can be flattened to produce a low-energy-spread accelerated beam. This work provided a robust theoretical framework for designing high-energy, high-efficiency, low-energy-spread PWFA stages.

These theoretical predictions have been validated by a series of landmark experiments. In 2014, an experiment at SLAC's FACET facility demonstrated beam loading with $\eta \approx 30\%$, confirming the nonlinear scaling laws \cite{Litos2014}. In 2018, an experiment at DESY demonstrated a transformer ratio greater than five using a tailored ramped driver beam, directly linking driver shaping to high-ratio acceleration \cite{Loisch2018}. More recently, Lindstrøm et al. demonstrated simultaneous high efficiency ($\eta > 30\%$) and low projected energy spread ($\approx 1\%$) by carefully matching the witness beam to the nonlinear wake, producing a beam quality suitable for driving a free-electron laser \cite{Linstrom2021}. The pursuit of even higher efficiencies and lower energy spreads remains a primary goal of ongoing research.

\subsection{Discovery of Beam-induced Plasma Self-Focusing Effect and Plasma Lens}

While the longitudinal electric field accelerates the particles, the transverse fields are equally consequential, as they govern the confinement and stability of the beams. A high-gradient accelerator is of little use if the beam cannot be kept focused and stable over the required distance. Remarkably, the plasma provides its own powerful focusing mechanisms, a feature central to the viability of PWFA.

Two primary sources of transverse focusing exist: the plasma wakefield itself, and a self-focusing force induced by the plasma's response to the beam's current.

In the linear regime, the transverse wakefield, $E_\perp$, is intrinsically linked to the longitudinal wakefield, $E_z$, through the Panofsky-Wenzel theorem \cite{Panofsky1956}, which implies $\nabla_\perp E_z = \partial E_\perp / \partial \zeta$. For an electron driver, the phase region suitable for accelerating a trailing electron bunch also provides a focusing force. As noted in early theoretical work \cite{Ruth1985}, this focusing force is highly linear near the propagation axis, a desirable quality for preserving beam emittance.

However, a distinct and often more powerful focusing mechanism arises from the interaction of the beam's current with the plasma: "relativistic self-focusing." In vacuum, the repulsive radial electric field ($E_\perp$) from a beam's space charge is almost perfectly canceled by the attractive Lorentz force from its self-generated magnetic field ($B_\theta$), leaving only a weak residual defocusing force proportional to $1/\gamma^2$. The presence of plasma dramatically alters this balance. The highly mobile plasma electrons rush in to neutralize the beam's space charge on a timescale of $\omega_p^{-1}$, effectively shielding the defocusing electric field $E_r$. The plasma response to the intruding beam also generates a return current that partially screens the beam's magnetic field, but this current neutralization is often incomplete within the beam volume. The result is that the beam's powerful, uncompensated magnetic field $B_\theta$ acts to pinch, or self-focus, the beam \cite{Chen1987, Chen1987Focusing,Su1990}.

\begin{figure}[h!]
    \centering
    \includegraphics[width=0.5\linewidth]{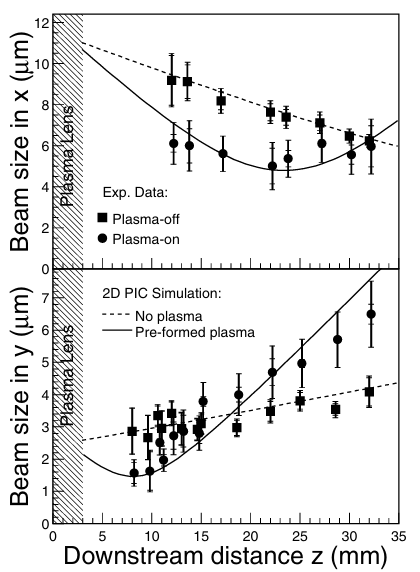}
    \caption{Measured beam envelope Gaussian widths in the horizontal (x) and vertical (y) planes for a 28.5 GeV positron beam, with (circle symbols) and without (square symbols) plasma focusing, as a function of downstream distance from a 3 mm nitrogen gas jet. The strong vertical pinch corresponds to an effective focusing gradient of $\sim4$ T/$\mu$m. Curves show particle-in-cell simulation results. Adapted from \cite{Ng2001}.}
    \label{fig:Positron_plasma_lens}
\end{figure}

A critical and unique feature of this self-focusing mechanism is its "charge-blind" nature. The focusing force arises from the interaction of a particle's velocity with the beam's collective magnetic field ($F_r \propto q v_z B_\theta$). Since both the particle charge $q$ and the magnetic field direction $B_\theta$ (which depends on the current direction) reverse for a positron compared to an electron, the resulting force $F_r$ remains radially inward for both species. This is in stark contrast to the transverse wakefield of a linear plasma wave, which is focusing for electrons but defocusing for positrons at accelerating phases. The prediction of a strong, simultaneous focusing for both electron and positron beams was profound and motivated the "plasma lens" concept \cite{Chen1987}. This effect was experimentally confirmed in the E-150 experiment at SLAC \cite{Chen1997SLAC}, which observed strong, symmetric focusing of both high-energy electron and positron beams in a plasma cell \cite{Chen1998Focusing,Ng2001}. In 2001, Ng et al. \cite{Ng2001} observed for the first time the positron beam self-focusing effect of a 28.5 GeV positron beam by a dense nitrogen plasma only 3 mm thick, achieving simultaneous reduction of the beam spot in both transverse dimensions. Figure~\ref{fig:Positron_plasma_lens} shows the measured beam envelopes with and without the plasma lens, together with corresponding PIC simulation results. The effective focusing strength reached $\sim4$ T/$\mu$m in the vertical plane, reducing the minimum beam cross-sectional area by a factor of about two, thus confirming the viability of plasma lenses for high-energy positron beams.

The highly nonlinear bubble regime can be seen as the ultimate manifestation of these principles. The ion channel, formed by the complete blowout of plasma electrons, provides a near-perfectly linear focusing force, $F_r \propto r$, arising from the uniform background of positive charge. This "ion-focused" regime provides an ideal transport channel for an electron witness bunch, capable of preserving emittance over long acceleration distances.

Despite these powerful focusing forces, beam stability remains a critical challenge. Misalignment between the witness beam and the ion channel can lead to the ``hose instability," a catastrophic transverse oscillation that can degrade emittance or even destroy the beam, as first analyzed by Whittum et al. \cite{Whittum1991PRL}. Early experiments at FACET observed the onset of hosing with misalignment as small as $10\,\mu\text{m}$ \cite{Adli2011IPAC}. More recent work has shown that this instability can be mitigated by introducing an energy spread or "chirp" in the witness beam. This creates a spread in the betatron oscillation frequencies along the bunch, which detunes the resonant growth of the instability, a mechanism known as BNS damping \cite{Mehrling2017,Mehrling2019PRAB}.

Another potential source of emittance degradation is ion motion. If the witness beam density is sufficiently high, the assumption of a static ion background breaks down. While early theoretical work predicted that ion motion could disrupt the linear focusing field \cite{Rosenzweig2005}, high-fidelity simulations by An et al. \cite{An2017PRL} showed that the effect is not catastrophic. The non-uniform collapse of ions introduces nonlinearities in the focusing force that in fact leads to controllable emittance growth.

Beyond mitigating instabilities for externally injected beams, significant effort has been directed towards generating high-quality witness beams in-situ. Schemes such as the ··Trojan Horse“ plasma cathode utilize a mixture of gas with different ionization thresholds \cite{Hidding2012PRL}. The driver ionizes the first species to form the plasma, while a precisely timed, focused laser pulse ionizes the second species directly within the bubble's accelerating and focusing fields. Simulations predict that this ionization-injection technique can generate beams with ultra-low emittance (on the order of nm-rad), meeting the stringent requirements for applications like free electron lasers.

Finally, accelerating positrons presents a unique and fundamental set of challenges, primarily because the wakefield phases that provide acceleration are inherently defocusing for positrons. This issue, critical for the development of a future e+e- collider, has led to a variety of innovative mitigation strategies, which will be discussed in detail in Section~\ref{sec:positronAcceleration}.

\section{From Concepts to Reality: Experimental Progress and Synergy}
The theoretical foundations laid in the mid-1980s opened the floodgates to a new field of research. The promise of GV/m accelerating gradients spurred a worldwide effort to turn these compelling concepts into physical reality. However, the path from theory to experiment is always shaped by the available technology. The landscape of the 1980s, particularly in laser technology, was vastly different from that of today. This technological context dictated the initial avenues of experimental exploration and set the stage for a fascinating early competition between two distinct approaches to plasma acceleration. This chapter traces the evolution of the field from these early days to the cutting-edge facilities of the present, culminating in a vision for their future synergy.

\subsection{The Early Landscape: Plasma Beat-Wave Accelerator}
In the wake of the 1979 Tajima-Dawson paper, the immediate challenge for the laser community was how to generate the required plasma wave. The original Laser Wakefield Accelerator (LWFA) concept envisioned using a single, extremely short laser pulse (on the order of a plasma period) to impulsively drive the wake. However, at the time, generating a laser pulse with the required terawatt-level power in such a short duration was technologically out of reach. It was not until the invention of Chirped Pulse Amplification (CPA) by Strickland and Mourou in 1985 \cite{Strickland1985} that a pathway to such ultrashort, ultra-intense pulses became clear. In this pre-CPA era, the community's efforts naturally converged on an alternative, resonant scheme that could work with the longer, less intense laser pulses that were available: the Plasma Beat-Wave Accelerator (PBWA).

In 1972, Rosenbluth and Liu~\cite{Rosenbluth1972} showed that the ponderomotive beating of two long, co-propagating laser pulses can resonantly drive an electron-plasma wave when the frequency difference $\Delta\omega=\omega_1-\omega_2$ matches the plasma frequency $\omega_p$.  Their original motivation was to explore laser-driven fusion, but the same mechanism soon attracted interest for wakefield particle acceleration.  The first systematic experimental study, *PBWA-I*, was reported by C.~Joshi and collaborators in 1982, who used two colinear CO\textsubscript{2} laser lines ($\lambda\simeq10.6$ and $9.6\:\mu$m) to generate GV\,m$^{-1}$ plasma waves confirmed by Thomson scattering~\cite{Joshi1982PBWA}.  Complementary particle-in-cell simulations presented at the same meeting by Sullivan and Godfrey predicted longitudinal fields ranging from 100 MeV cm$^{-1}$ to 10 GeV cm$^{-1}$ and argued that staging could, in principle, reach TeV beam energies~\cite{Sullivan1982}.  

Direct evidence of relativistic-amplitude beat-wave excitation followed in 1985: Clayton \textit{et al.} measured density perturbations $\delta n/n\sim 2\%$ and inferred sub-GV/m fields in a Hydrogen plasma~\cite{Clayton1985}.  In the same year, Ebrahim \textit{et al.} demonstrated that $\sim150$ keV electrons produced by stimulated Raman backscattering were trapped and accelerated to $3$ MeV over a $0.5$ mm interaction length— an effective gradient of $\sim6$ GV\,m$^{-1}$~\cite{Ebrahim1985}.  Throughout the late-1980s and early-1990s, this plasma-beat-wave accelerator (PBWA) concept remained a central theme in laser-plasma acceleration research.

It was in this landscape that PWFA emerged as a competing paradigm. As reviewed at the 1986 SLAC Summer School by Ruth and Chen \cite{Ruth1986}, the field in the mid-80s was thus characterized by a side-by-side comparison of these two distinct approaches. A comparative view of their perceived advantages and challenges at the time is instructive:

\begin{description}
	\item[The Driver] The PBWA relied on high-power gas lasers (typically CO$_2$ lasers), which were complex systems. The PWFA, in contrast, proposed to use relativistic electron beams from conventional linacs, which were a mature, well-understood technology, particularly at major high-energy physics laboratories like SLAC. This was seen as a significant advantage in terms of driver availability and stability.
	
	\item[Resonance Condition] The PBWA was critically dependent on a fine-tuned resonance ($\Delta\omega = \omega_p$), which required precise matching condition of the plasma and laser. To obtain longer accelerating distance, well-controlled plasma density is necessary. Any deviation from the resonant density would lead to phase slippage and detuning, limiting the effective interaction length. The PWFA, which impulsively drives the plasma wake rather than relying on a resonant build-up, was perceived as being far more robust against small variations in plasma density.
	
	\item[Wave Amplitude and Control] In PBWA, the final wave amplitude was limited by relativistic detuning, where the nonlinear shift in the plasma frequency eventually breaks the resonance condition. The PWFA, particularly in the nonlinear regime, was predicted to be capable of reaching the wave-breaking limit more directly, with the ultimate gradient controlled simply by the driver's charge density.

    \item[Efficiency and Scalability] The consequences of the different driving mechanisms were quantified by Ruth and Chen. For a given accelerating gradient, the PWFA was shown to have a significantly higher efficiency in transferring energy from the driver to the plasma wake. More critically, due to the diffraction limits of lasers, scaling the PBWA to longer acceleration stages required larger laser spot sizes, which in turn demanded exponentially higher laser power. The PWFA, using particle beams that do not suffer from such diffraction, appeared far more scalable to the long distances required for a high-energy collider.
	
	\item[Overall Complexity] The PWFA appeared, at least conceptually, to be a simpler system: inject a single, pre-existing beam into a plasma source. The PBWA required the complex alignment and synchronization of two high-power laser beams, an electron injector, and a uniform plasma target.
\end{description}

In summary, the side-by-side comparison of the era, as articulated by Ruth and Chen, presented the PWFA as a promising alternative that leveraged the mature infrastructure of accelerator laboratories and appeared more robust, efficient, and scalable than the resonant PBWA scheme \cite{Ruth1986}. This perception fueled the first generation of PWFA experiments designed to demonstrate the core principles of beam-driven wakefield excitation.

By the mid-1990s, with the revolutionary impact of Chirped Pulse Amplification now fully realised, the field had matured significantly. The landscape of plasma-based concepts—including the original short-pulse LWFA, the self-modulated LWFA, and the rapidly advancing PWFA—was systematically captured in a landmark 1996 review article by Esarey, Sprangle, Krall, and Ting \cite{Esarey1996Overview}. This comprehensive overview provided the first unified theoretical framework for all major plasma accelerator schemes, solidifying their status as a credible path towards the next generation of high-energy accelerators. It is from this era that the modern, distinct experimental frontiers of PWFA and short-pulse LWFA, which we will explore next, truly began to flourish.

\subsection{Frontiers of Beam-Driven Acceleration (PWFA)}

\begin{figure}[h!]
    \centering
    \includegraphics[width=1.0\linewidth]{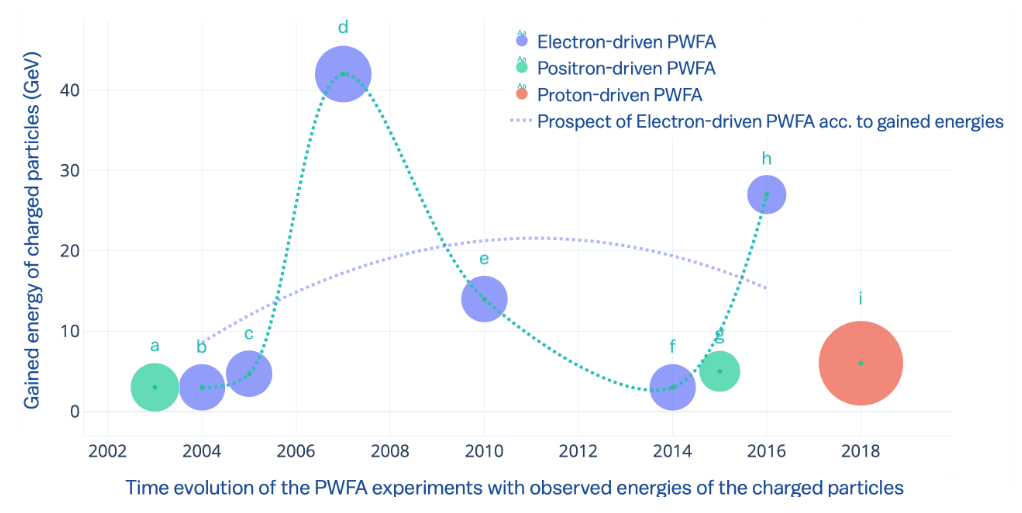}
    \caption{Time evolution of plasma wakefield acceleration (PWFA) experiments showing the gained energies of accelerated particles for electron-driven, positron-driven, and proton-driven schemes. Data points correspond to landmark experiments, with circle size indicating the driver beam energy. The dotted green line connects reported experimental results, while the purple dashed line indicates a projected trend for electron-driven PWFA energy gain. Adapted from \cite{Cakir2020}.}
    \label{fig:PWFA_gained_Energy}
\end{figure}

Fueled by its conceptual robustness and the availability of high-quality electron beams, the PWFA program progressed from early proof-of-principle tests to a series of groundbreaking experiments at large-scale facilities. These experiments have not only validated the core theoretical predictions but have also pushed the boundaries of energy gain, efficiency, and beam quality, demonstrating the potential of PWFA as a technology for future colliders and light sources. Figure~\ref{fig:PWFA_gained_Energy} provides a chronological overview of the maximum energy gains achieved in major PWFA experiments over the past two decades, categorized by driver types. The clear progression—from early few-GeV demonstrations to multi-tens-of-GeV gains—highlights the rapid advancement of beam-driven plasma acceleration. Notably, electron-driven schemes (blue) have reached an energy gain beyond 40 GeV in single-stage acceleration, while recent positron-driven (purple) and proton-driven (red) experiments have demonstrated multi-GeV capability. This overview sets the stage for the discussion of key facility programs in the following subsections. 

\subsubsection{SLAC: Pioneering High-Gradient, High-Efficiency Acceleration}

The Stanford Linear Accelerator Center (SLAC), with its unique high-energy, high-current electron and positron beams, has been the primary engine of PWFA research for decades. While the first observation of plasma wakefield acceleration occurred at the Argonne National Laboratory \cite{Rosenzweig1988a}, the field truly matured at SLAC through a strategically phased series of experiments culminating in the FACET and now FACET-II user facilities.

The initial experiments at the Final Focus Test Beam (FFTB) facility systematically built the foundations. The E-150 experiment successfully validated the concept of a thin plasma lens, showing strong, symmetric focusing for electron beams and, for the first time, the observation of plasma focusing of positron beams—a critical path for future linear colliders \cite{Chen1987Focusing,Chen1997SLAC}. This was followed by the E-157 experiment, which achieved a $\sim 0.5\,\text{GV/m}$ acceleration gradient and accelerated particles up to 1 GeV over a meter-scale lithium plasma source. The landmark E-167 experiment then provided the first definitive evidence of ``energy doubling", where the tail of a single 42 GeV electron bunch was accelerated by the wake generated by all beam particles ahead of the tail, with some particles gaining over 42 GeV and experiencing gradients of $\approx 52\,\text{GV/m}$ \cite{Hogan2005}.

The next generation facility at SLAC, the Facility for Advanced Accelerator Experimental Tests (FACET), was designed to explore the highly nonlinear blowout regime using a "two-bunch" (driver-witness) configuration. In the E-200 experiment, a witness bunch was efficiently loaded into the wake, resulting in a 9 GeV average energy gain over 1.3 meters with a remarkable 30\% driver-to-witness energy transfer efficiency and $\sim 5\%$ energy spread \cite{Litos2014}. This result ushered in the era of high-efficiency, high-gain PWFA. The study of positron acceleration was also a key focus; the E-200 experiment demonstrated the first multi-GeV energy gain for a positron beam in a self-loaded plasma wake, a significant step forward \cite{Corde2015}. This result, along with other approaches to address the challenge facing positron acceleration in the nonlinear regime, is further explored in Section \ref{sec:positronAcceleration}.

Building on the success of FACET, the current FACET-II facility is now commissioning a suite of critical campaigns. The baseline experiment (E-300) aims to push efficiency beyond 40\% while preserving emittance. Other projects tackle key physics issues such as the suppression of hosing instability (E-302) and direct optical visualization of the bubble structure (E-324). With its unique beams, flexible user-facility model, and a program that has progressively demonstrated all essential ingredients—high gradient, high efficiency, beam quality preservation, and positron compatibility—SLAC remains central to the PWFA roadmap, particularly for a future PWFA-based $e^+e^-$ collider \cite{Cao2024positron}.

\subsubsection{CERN (AWAKE): PWFA Driven by Proton Beams}

While most PWFA research employs electron drivers, the AWAKE (Advanced Wakefield Experiment) at CERN pioneered a new version of PWFA: taking advantage of the 400 GeV proton beams available from the in-house Super Proton Synchrotron (SPS) as the driver~\cite{Caldwell2009}. A single SPS bunch carries $\sim 19\text{kJ}$ of energy, about two to three orders of magnitude more than typical electron drivers. This immense energy budget is, in principle, sufficient to accelerate a witness beam to TeV-scale energies within a single, continuous plasma stage, a prospect unique to the proton-driven paradigm.

As the bunch length ($\sigma_z \approx 6\text{ cm}$) of SPS proton beams is much longer than the plasma wavelength ($\lambda_p \approx 1\text{ mm}$), direct resonant driving is impossible. AWAKE masterfully exploits the self-modulation (SM) instability~\cite{Kumar2010}, whereby a long proton bunch propagating in plasma naturally splits into a train of micro-bunches spaced by $\lambda_p$. These micro-bunches then resonantly excite strong wakefields, echoing the original bunch-train scenario of Chen et al. \cite{Chen1985}. Run 1 (2016–2018) successfully demonstrated this principle, seeding the instability with a laser ionization front and accelerating externally injected electrons from 19 MeV to 2 GeV in a 10-meter plasma source \cite{Adli2018}.

Run 2 is now unfolding in carefully planned phases \cite{Gschwendtner2022, Gschwendtner2024}. Run 2a successfully demonstrated seeding the SM with a preceding electron bunch, achieving sub-picosecond phase stability. Run 2b is optimizing the wakefield amplitude using tailored plasma density profiles. The forthcoming Run 2c aims to inject a 150 MeV witness beam and accelerate it to $\sim 10\,\text{GeV}$ with excellent beam quality, targeting gradients of 0.5-1 GV/m. The final phase, Run 2d, will test scalable plasma sources (discharge or helicon) for a future 100-meter-scale experiment, paving the way for applications like fixed-target dark-sector searches or a very-high-energy electron-proton collider.

\subsubsection{DESY (FLASHForward) and INFN (SPARC\_LAB): Towards Applications}

Beyond the high-energy frontier, a major focus of PWFA research is producing beams with the quality required for applications like free-electron lasers (FELs). This requires not only high energy beams but also low energy spread and, critically, preservation of the beam's ultra-low emittance.

The FLASHForward experiment at DESY (Hamburg) leads in this area, utilizing the high-quality electron beams from the FLASH superconducting linac. After demonstrating high efficiency and low energy spread in its commissioning phase, the team recently achieved a landmark result: the first demonstration of a brightness-preserving PWFA stage, maintaining a 2.8 mm-mrad emittance over a 50 mm plasma cell \cite{Lindstrom2024}. This achievement is of profound importance, as emittance preservation is arguably the most critical figure of merit for any application-oriented advanced accelerator, especially for driving FELs or for injection into subsequent stages. The complex mechanisms governing emittance growth, such as beam-plasma instabilities and beam mismatch, and the various mitigation techniques are systematically analyzed in the comprehensive review by Lindstrøm \& Thévenet \cite{Lindstrøm2022Preservation}. Having now realized all three hallmarks—GV/m gradient, $\ge 40\%$ efficiency, and full beam-quality preservation—the team is focused on scaling the energy gain while pushing towards MHz-repetition-rate-compatible plasma technologies.

Similarly, the SPARC\_LAB facility at INFN-LNF (Frascati) has become a hub of innovation, with a program that feeds directly into the forthcoming EuPRAXIA@SPARC\_LAB user facility. They have systematically studied resonant PWFA, using "comb-beam" trains of up to N=5 sub-picosecond bunches to demonstrate the N-fold field enhancement, reaching fields of $\approx 3\,\text{GV/m}$ \cite{Bosco2024}. A landmark achievement was the first demonstration of FEL lasing driven by a witness beam from a compact PWFA module, achieving saturation at 60 nm \cite{Pompili2022}. This proof-of-principle experiment underpins the EuPRAXIA project, a $\sim 160$-meter facility that will couple a 1 GeV linac to a 3-5 GeV PWFA booster to drive a soft X-ray FEL, targeting 100 Hz user operation. Together, these application-oriented programs are critical for translating the promise of plasma acceleration into tangible scientific tools.

\subsection{New Chapters of Laser-Driven Acceleration (LWFA)}
While PWFA research leveraged the existing infrastructure of large accelerator laboratories, the LWFA field was fundamentally revolutionized by a breakthrough in laser science: the invention of Chirped Pulse Amplification (CPA) by Strickland and Mourou in 1985 \cite{Strickland1985}. CPA made it possible to generate ultrashort (femtosecond-scale) laser pulses with unprecedented peak powers (terawatt to petawatt levels). This technology directly unlocks the original impulsive driving mechanism envisioned by Tajima and Dawson \cite{Tajima1979}, thereby transforming LWFA from a challenging concept into a vibrant, rapidly evolving, and prevalent field of research.

The modern LWFA operates in the highly nonlinear``bubble" or ``blowout" regime, analogous to its beam-driven counterpart. An intense, ultrashort laser pulse characterized by the normalized vector potential $a_0 = eA_0/m_e c^2 \gtrsim2$, where $A_0$ is the laser vector potential, violently expels plasma electrons by the ponderomotive force, forming a bubble-like wake structure that can trap and accelerate electrons to enormous energies. The field has seen a series of landmark achievements that have progressively pushed the energy frontier. A trio of papers in 2004 from independent groups at LBNL, Imperial College London, and LOA (France) demonstrated the generation of quasi-monoenergetic electron beams in the 100-MeV range, a critical step that showed LWFAs could produce not just high energy, but also high-quality, beams \cite{Geddes2004, Mangles2004, Faure2004}. 

\begin{figure}[h!]
    \centering
    \includegraphics[width=0.8\linewidth]{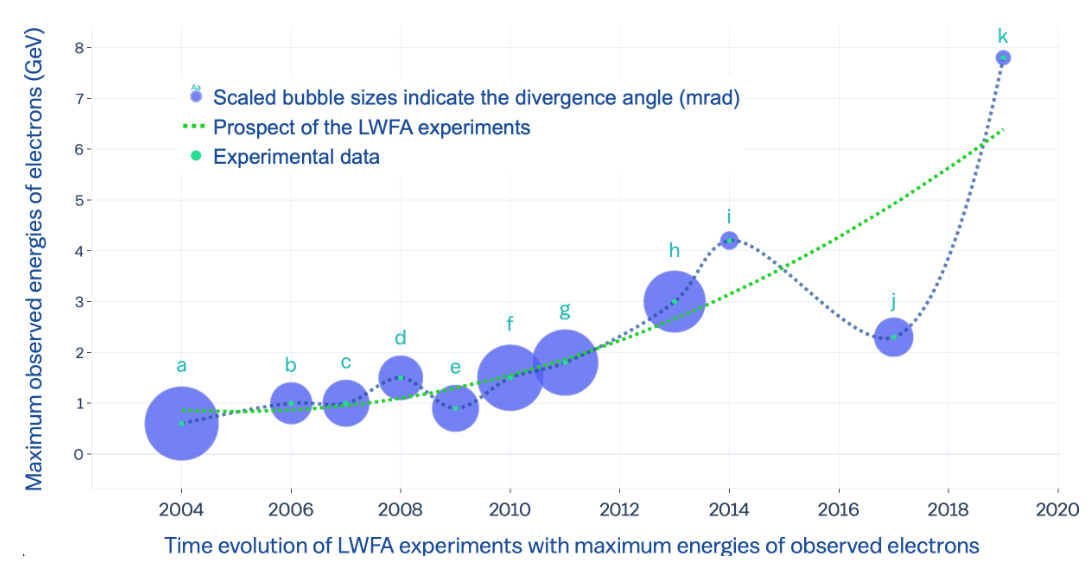} 
    \caption{Time evolution of the maximum electron energy from landmark LWFA experiments between 2004 and 2020. The plot illustrates the rapid ascent to the multi-GeV frontier, with data point sizes scaled to the reported beam divergence (mrad). As discussed in the text, progress has continued beyond this timeframe, with energies exceeding 10~GeV achieved since 2024. Adapted from~\cite{Cakir2020}.}
    \label{fig:lwfa_progress}
\end{figure}

This was followed by the "GeV era," where careful control of the laser-plasma interaction over longer distances, often using pre-formed plasma channels, enabled the acceleration of electrons to energies exceeding 1 GeV in a single, centimeter-scale stage \cite{Leemans2006}. More recently, by guiding a petawatt-class laser pulse through a 20-cm-long plasma channel, the BELLA Center at LBNL has achieved a record energy of nearly 8 GeV, a testament to the remarkable gradients and sustained acceleration possible with LWFAs \cite{Gonsalves2019}. The progress is visually summarized in Fig.~\ref{fig:lwfa_progress}, which chronicles the evolution of maximum electron energies achieved in LWFA experiments. The data, compiled up to 2020~\cite{Cakir2020}, clearly illustrates the rapid ascent from the 100-MeV scale to the multi-GeV frontier. It is important to note that this trend is continuing, with subsequent breakthroughs pushing the energy frontier even further.

In 2024, 10 GeV electrons were produced at the Texas Petawatt Laser Facility using nanoparticle-assisted injection \cite{Constantin2024}. Complementary channel-guided work at BELLA later that year reached $9.2 GeV$ with sub-percent energy spread \cite{picksley2024}. Most recently, Rochafellow et al. have reported $\ge30\%$ laser-to-electron efficiency at the $10$ GeV level \cite{Rockafellow2025}.

To maintain the intense laser focus required for acceleration over many Rayleigh lengths, several guiding techniques have been developed. These include using pre-formed plasma channels created by hydrodynamic expansion of a laser-ignited spark in a gas jet, or using gas-filled capillary discharge waveguides. These techniques have been essential for reaching the multi-GeV energy scale.

The progress in LWFA has been extensively documented in several excellent review articles, which provide a comprehensive overview of the underlying physics, technologies, and experimental results \cite{Esarey2009, Hooker2013, Leemans2019}. The current frontiers of LWFA research are focused on several key challenges, each with its own recent landmark achievements that define the path forward.

\subsubsection{Staging for Higher Energies}
To reach energies relevant for a high-energy collider (TeV scale), cascading multiple LWFA stages is essential. The advent of multi-petawatt laser systems allows for larger laser spot sizes to reach the required intensities, thereby significantly increasing the Rayleigh length and mitigating the need for guiding over short distances. However, even if the laser focus can be maintained indefinitely, two fundamental limit persists: dephasing and pump depletion length. The first limitation results from the fact that the accelerated electrons travel at nearly the speed of light, they inevitably outrun the accelerating phase of the plasma wave, which propagates at the laser's typically lower group velocity in plasma. This dephasing length imposes an intrinsic limit on the maximum energy gain achievable in any single, uniform plasma stage, irrespective of the available laser power; Due to the nonlinear interaction between the driver (pump) and the plasma, the strong depletion of the pump also sets a maximum acceleration length. To overcome this issue, staging is the only viable path to the highest energies. This requires not only generating a high-quality beam in the first stage but also capturing, transporting, and injecting it into a subsequent stage with minimal degradation.

A significant challenge is the transport optics, as conventional magnetic quadrupoles are too weak and bulky. A promising solution is the use of active plasma lenses, which can provide kT/m focusing gradients \cite{vanTilborg2015}. Furthermore, managing the laser driver between stages is critical. A notable recent achievement demonstrated the guiding of a high-power laser using a transition curved plasma channel \cite{Luo2018,Zhu2023}, which is an encouraging step toward laser refreshing without disturbing the electron beam line.

\subsubsection{Towards High-Repetition-Rate and High-Efficiency Operation}
Most record-breaking LWFA experiments have been performed with low-repetition-rate ($\sim 1$ Hz) high-power Ti:Sapphire laser systems. For many applications, from fundamental science to medicine and industry, migrating to high-repetition-rate (kHz or higher) operation is paramount. This transition is being driven by rapid advancements in laser technology, particularly high-average-power systems like Yb-doped lasers and coherent combination of fiber lasers. Demonstrations of stable, tens of MeV-class electron beam generation at kHz repetition rates represent a major step in this direction \cite{Salehi2021}. In addition, improvement of wall-plug-to-beam efficiency, which is currently very low ($<0.1\%$), is a critical long-term goal. This involves optimization of both the laser system and the laser-plasma energy coupling.

\subsubsection{Beam Quality and Control for Precision Applications}
While tremendous progress has been made in energy gain, achieving simultaneous control over all beam parameters—energy spread, emittance, pointing stability, and charge—remains a central challenge. The self-injection process that often generates the beam can be inherently unstable. To overcome this, advanced injection techniques have been developed to decouple the beam injection from the acceleration process. These include using sharp density transitions (density-downramp injection) \cite{Geddes2008}, colliding laser pulses\cite{Faure2006}, nano-particle assisted \cite{Constantin2024}, to controllable trigger injections. These methods have demonstrated the ability to produce beams with percent-level energy spread and improved stability, which are crucial for applications.

\subsubsection{Driving Applications: Compact Light Sources}
Perhaps the most tangible near-term application of LWFA is the creation of compact, university-scale radiation sources. One of the most significant milestones in this area has been the demonstration of an LWFA-driven free electron laser (FEL). The first lasing of an FEL driven by a plasma accelerator was achieved in the infrared, but a landmark result from the Shanghai Institute of Optics and Fine Mechanics (SIOM) in 2021 demonstrated high-gain amplification in the extreme ultraviolet (EUV) range, reaching saturation and producing an intense, coherent pulse \cite{Wang2021}. This was a pivotal moment, proving that LWFA beams possess the high brightness (high peak current and low emittance) required for FEL applications. Beyond FELs, the relativistic electron beams traversing the plasma bubble naturally produce intense, incoherent X-ray beams via betatron oscillations \cite{Rousse2004}, where the energy is now reaching multi-$\mu\text{J}$ \cite{Kneip2010Betatron}. These demonstrated that applications are rapidly maturing LWFA from a scientific curiosity into a powerful and compact scientific tool.

The idea of using a relativistic \emph{plasma wave} as a purely electric wiggler for a free-electron laser (FEL) was pioneered at UCLA under J.M.~Dawson. Yan and Dawson introduced the concept of an "ac FEL" in 1986\cite{YanDawson1986}, and in 1987, Joshi \emph{et al.} analyzed the plasma-wave wigglers, showing the formal equivalence of magnetic, electromagnetic and purely electric (plasma) wigglers while also highlighting plasma-imposed gain limits and dephasing constraints~\cite{Joshi1987_JQE,Joshi1987_PAC}. Although plasma wigglers can reach very short effective periods and large strength parameters $K$, experimental demonstrations to date have produced spontaneous x-rays via betatron motion in ion channels rather than coherent FEL gain~\cite{Wang2002_PRL}. Recent proposals aim to narrow the radiation bandwidth and extend the effective number of undulator periods using plasma-channel undulators driven by high-order laser modes with phase locking and tapering; these studies identify bandwidth uniformity, period count, and stability as the principal hurdles for FEL lasing~\cite{Rykovanov2016_PRAB,Wang2017_SciRep}.

The remarkable progress of LWFA, powered by the parallel advancement of laser technology, has established it as a leading candidate for developing compact, university-scale accelerators and has created a rich and synergistic research landscape alongside the beam-driven PWFA.

\subsection{The Positron Conundrum: A Critical Challenge for Colliders}
\label{sec:positronAcceleration}

While both PWFA and LWFA have demonstrated remarkable success in accelerating electrons to multi-GeV energies, the path toward a high-energy electron-positron (e$^+$e$^-$) collider is contingent on solving a fundamental challenge: the stable acceleration of positrons. The difficulty arises specifically in the high-gradient, nonlinear blowout regime. This challenge is rooted in the intrinsic mass asymmetry of a conventional plasma: the wake structure is formed by highly mobile electrons being expelled from a quasi-static background of heavy ions. The resulting ion column, or "bubble," creates a near-perfect, linearly focusing channel for electrons but is, for the same reason, strongly defocusing for positrons.

It is worth noting that this is a nonissue in the linear perturbation regime, where, as discussed in Sec.~\ref{sec:linear_regime}, phase regions that are simultaneously accelerating and focusing exist for both electrons and positrons. However, the significantly higher accelerating gradients achievable in the blowout regime make it the primary focus for future high-energy applications. The quest for a viable positron solution in the high-gradient blowout regime is therefore a critical task for the PWFA community, and represents a fundamental challenge that must also be addressed for any future LWFA-based collider.

Overcoming this "positron conundrum" has become a central and vibrant area of plasma accelerator research, leading to several promising mitigation strategies:

\begin{itemize}
    \item \textbf{Self-Loading and Beam Shaping}: As demonstrated in a landmark experiment at SLAC's FACET facility, a positron beam can be accelerated by multi-GeV energies if the beam itself is intense enough to significantly modify the wake, a process known as self-loading~\cite{Corde2015}. By carefully shaping the positron driver bunch (e.g., with a step-shaped profile) and the witness bunch, it is possible to create a wake structure that is simultaneously accelerating and focusing, as is being explored in the FACET-II E333 experiment~\cite{Zhou2022PRAB, Lee2023AAC}.
    
    \item \textbf{Hollow Plasma Channels}: One of the leading concepts is to use a plasma with a hollow channel at its center~\cite{Gessner2016NatComm}. In such a structure, the transverse wakefields within the channel can be engineered to be focusing for positrons in the accelerating phase, providing a stable transport channel. The challenge lies in creating and maintaining these precisely shaped plasma profiles over long distances.    

    \item \textbf{Hybrid Beam Schemes}: Another innovative approach involves using an ``escort'' electron beam that co-propagates with the positron beam. The space-charge field of this electron filament can locally cancel the plasma's defocusing force, creating a net focusing force for the positrons.

\end{itemize}

The continued development and demonstration of a robust and efficient positron acceleration scheme represents a critical milestone on the global roadmap toward a plasma-based collider. The diversity of the solutions being pursued, from intricate beam shaping to novel plasma structures, underscores the field's ingenuity and highlights this area as a frontier of active research and development. For a comprehensive overview of the current status, we refer the reader to the recent review by Cao et al.~\cite{Cao2024positron}.

Beyond the schemes discussed above, new concepts continue to emerge. One promising direction involves leveraging positron-driven wakes in novel ways. While using a positron bunch to drive the wake naturally attracts plasma electrons to form potentially useful accelerating and focusing regions, pushing to the high-gradient nonlinear regime ($\phi_0 \gtrsim 1$) is challenging. The attracted electrons tend to collapse onto the axis, forming a sharp density cusp that creates highly nonlinear transverse fields, unsuitable for preserving beam emittance.

\begin{figure}[h!]
    \centering
    \includegraphics[width=\linewidth]{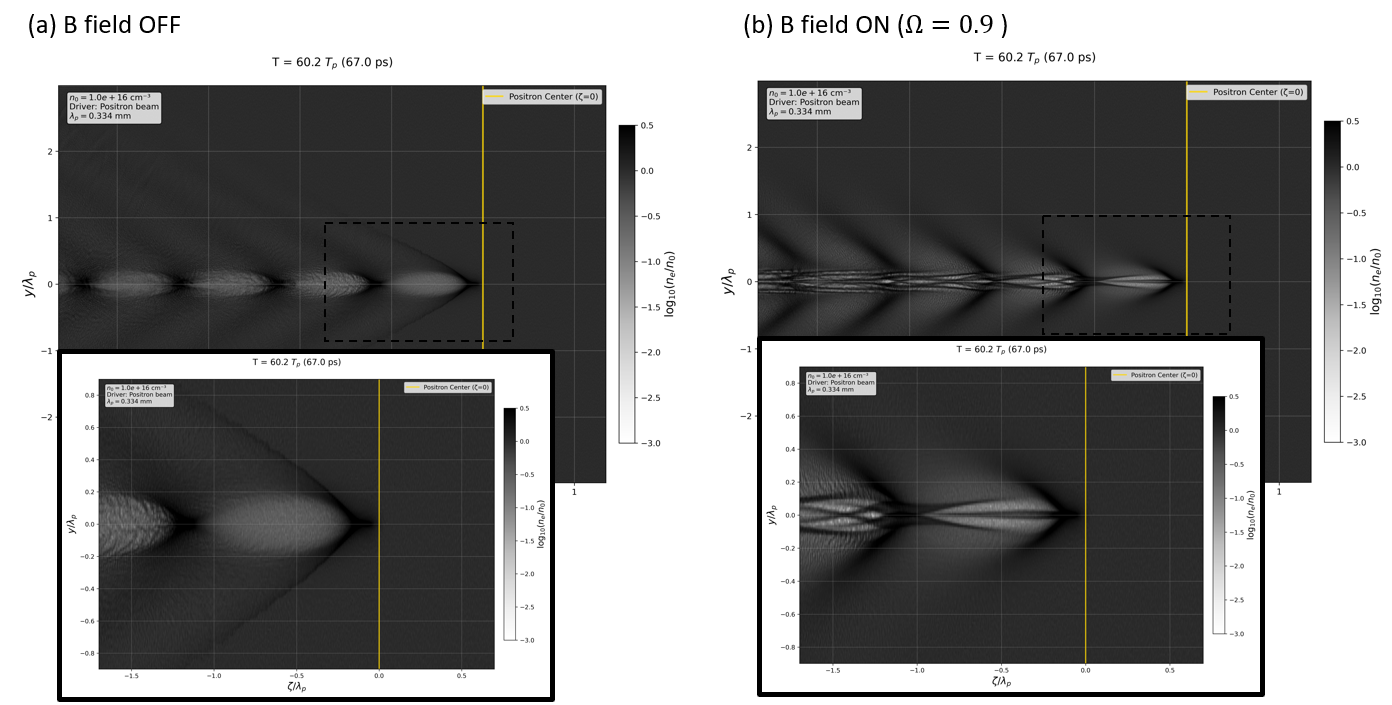}
    \caption{PIC simulation of plasma electron density ($\log_{10}(n_e/n_0)$) in a wake driven by a positron driver ($\phi_0=2.5$). (a) Without an external magnetic field, a full electron blowout is observed. (b) With a longitudinal magnetic field ($\Omega=0.9$), a stable electron column is formed on-axis, confined within the larger bubble structure. The insets show a magnified view of the region of interest.}
    \label{fig:positronAcc_electron_density}
\end{figure}

As a potential solution, we propose the application of a strong, external longitudinal magnetic field ($B_z$) to regularize this electron behavior. We have investigated this concept via 2D3V particle-in-cell (PIC) simulations using the EPOCH code\cite{EPOCH}. The field's influence is quantified by $\Omega\equiv \omega_c/\omega_p$, the ratio of the electron cyclotron frequency ($\omega_c$) to the plasma frequency. Figure~\ref{fig:positronAcc_electron_density} compares the plasma electron density driven by a positron bunch ($\phi_0=2.5$) with and without this field. In the unmagnetized case (Fig.~\ref{fig:positronAcc_electron_density}a), a complete electron blowout occurs. However, with a strong field of $\Omega=0.9$ (corresponding to $B_z \approx 29$~T), the electrons are magnetically confined, preventing blowout and cusp formation, and establishing a stable on-axis electron column (Fig.~\ref{fig:positronAcc_electron_density}b).

\begin{figure}[h!]
    \centering
    \includegraphics[width=\linewidth]{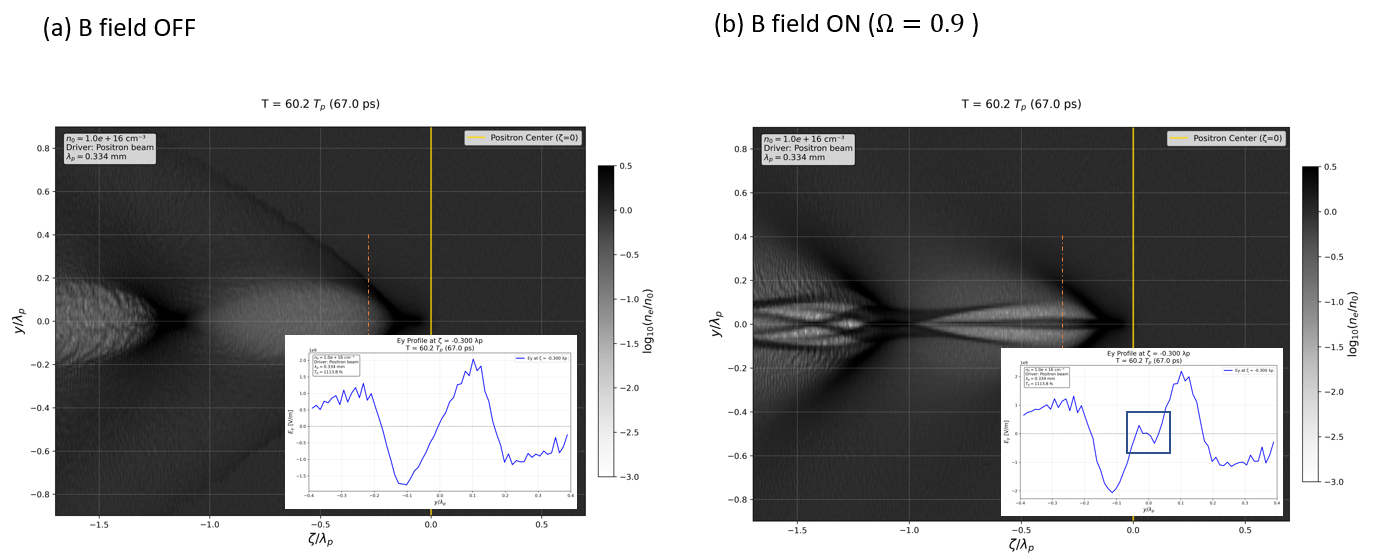}
    \caption{Field structure and accelerating phase in the positron-driven wake. (a) In the unmagnetized case, the wake is purely defocusing for positrons. (b) In the magnetized case, the on-axis electron column creates a focusing force. The inset shows the longitudinal accelerating field ($E_z$) on-axis, with the blue box highlighting a phase-stable region where both acceleration ($E_z < 0$) and transverse focusing coexist.}
    \label{fig:positronACC_field_structure}
\end{figure}

This magnetically-guided structure fundamentally reshapes the wake's transverse fields. As shown in Fig.~\ref{fig:positronACC_field_structure}, the on-axis electron column partially neutralizes the ion charge, creating a focusing force for positrons. This allows a trailing witness bunch to be stably trapped in a region that is also accelerating ($E_z < 0$), a condition highlighted by the blue box in Fig.~\ref{fig:positronACC_field_structure}(b). Such a phase-stable region is absent in the unmagnetized case. This approach is conceptually analogous to the ``electron-escort'' scheme, but it achieves the guiding effect without requiring a separate escort beam, relying instead on an intense magnetic field potentially available from state-of-the-art superconducting magnets. This demonstrates that there is still a rich landscape of physics to explore in the quest for an optimal positron acceleration solution.

\subsection{A Synergy: Hybrid Laser-Plasma Wakefield Accelerator (H-LPWFA)}
The parallel development of beam-driven and laser-driven plasma accelerators has created a rich and competitive research landscape. Yet, it has also revealed a fundamental tension: PWFA offers high energy-transfer efficiency and leverages mature accelerator technology but is often tied to large, costly conventional injectors; conversely, LWFA, powered by increasingly compact high-power lasers, offers a path towards university-scale accelerators but faces challenges in efficiency and shot-to-shot stability. A natural and elegant path forward, therefore, is to merge the two paradigms into a \emph{hybrid wakefield accelerator}, a concept that aims to harness the respective strengths of both schemes \cite{Hidding2010}.

The core idea of the hybrid accelerator is to use an all-optical, laser-driven setup to create a high-quality electron beam, which then immediately serves as the driver for a subsequent, high-efficiency PWFA stage. This "bootstrapping" approach unfolds in two distinct steps:
\begin{enumerate}
    \item \textbf{LWFA for the Injection Stage:} An intense laser pulse interacts with a gas target to generate a stable, low-emittance, relativistic electron bunch via a controlled injection mechanism (e.g., density-downramp or ionization injection). The primary goal of this stage is to produce a high current driver beam, not necessarily to achieve maximum energy gain.
    \item \textbf{PWFA for the Acceleration Stage:} This laser-generated electron beam is then transported a short distance and injected into a second plasma cell. Here, it acts as a PWFA driver for a stage optimized for high transformer ratio and high efficiency, accelerating a witness bunch to high energies.
\end{enumerate}

This hybrid architecture offers several profound advantages. First, it decouples the electron beam generation from the main acceleration process, allowing each stage to be optimized independently. This effectively creates a compact, all-optical PWFA, replacing a kilometer-scale conventional front-end with a meter-scale laser system. Second, electron beams generated by LWFAs are intrinsically ultrashort (femtosecond-scale) with extremely high peak currents (kiloampere-level). Such beams are ideal drivers for a PWFA, naturally suited for generating high-transformer-ratio wakes without the complex bunch compression required in conventional accelerators. Indeed, recent experiments have already successfully used such laser-generated, kA-level beams to drive high-gradient PWFA stages, confirming the viability of this crucial first step \cite{Kurz2021, Foerster2022}.

The promise of this synergistic approach has spurred a global research effort, with theoretical and computational studies mapping out viable parameter sets and exploring the physics of inter-stage transport and beam matching. The concept has been independently proposed and investigated by researchers worldwide, with significant contributions from groups in Europe and Asia who are exploring staged and hybrid designs for future compact light sources and colliders \cite{Hidding2023Review, Chang2025}.

Experimentally, demonstrating a fully integrated and optimized hybrid accelerator represents a convergence of the PWFA and LWFA communities, leveraging expertise from both sides. The successful realization of such a scheme would be more than a technical achievement; it would represent a paradigm shift in accelerator design, paving the way for compact, high-repetition-rate, high-energy machines that were previously inconceivable. It is a testament to the field's maturity that it is evolving from competing approaches towards a unified, synergistic vision for the future.

\section{Laboratory Astrophysics: Probing Cosmic Extremes with Lasers, Beams, and Plasmas}
\subsection{The New Symbiosis: Lasers and Beams as a Bridge to the Cosmos}
Having established the principles and rapid experimental progress of plasma wakefield accelerators in the preceding sections, we now pivot from their application as terrestrial particle accelerators to their profound potential as tools for emulating the cosmos itself. The last few decades have witnessed concurrent revolutions: our understanding of the universe has deepened, revealing phenomena of unimaginable energy and scale, while laser technology has achieved intensities that can recreate the extreme conditions found in astrophysical environments. This convergence has given rise to the vibrant field of \emph{laboratory astrophysics}, where high-intensity lasers, particle beams, and plasmas serve as a bridge, allowing us to probe the fundamental physics of the universe on a tabletop \cite{Remington1999, drake2010,Chen2014laser}.

\begin{figure}[h!]
    \centering
    \includegraphics[width=0.7\linewidth]{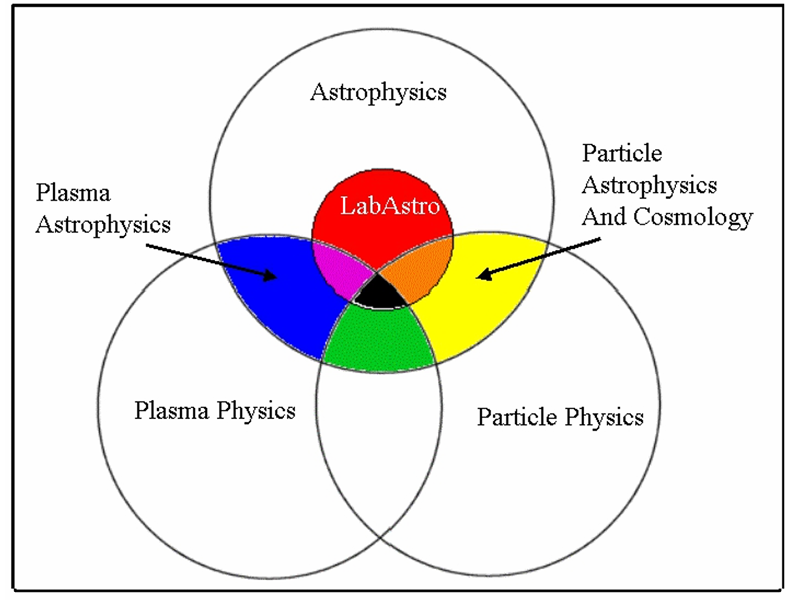}
    \caption{A Venn diagram illustrating the interdisciplinary nature of laboratory astrophysics, situated at the confluence of particle physics, astrophysics, and plasma physics. High-intensity lasers are a key enabling technology for this field. Adapted from \cite{Chen2014laser}}
    \label{fig:venn_diagram}
\end{figure}

As illustrated in Fig.~\ref{fig:venn_diagram}, laboratory astrophysics resides at the intersection of plasma physics, particle physics, and astrophysics. It provides a unique methodology to address frontier questions in cosmology and particle astrophysics that are otherwise inaccessible. These cosmic phenomena typically involve one or more of the following conditions: 1) extremely high-energy particle events; 2) ultra-high-density, high-temperature processes; and 3) super-strong field environments. By creating micro-analogs of these conditions in the laboratory, we can directly investigate the underlying dynamics, benchmark theoretical models, and even discover new physical principles that govern extreme environments. While the field of laboratory astrophysics is broad, encompassing critical studies of phenomena such as magnetic reconnection, hydrodynamic instabilities, and stellar opacities \cite{yamada2010, remington2006}, this review will focus on two examples that are deeply intertwined with the physics of plasma wakefields and relativistic acceleration. 

First, we will examine how the physics of plasma wakefield acceleration, our central topic, provides a compelling new mechanism to explain the origin of the most energetic particles in the universe. Second, we will introduce the concept of Unruh radiation, a fundamental prediction of quantum field theory in curved spacetime, and discuss how accelerated plasma systems can serve as a platform to probe its existence. This will set the stage for Section \ref{sec:AnaBHEL}, where we will detail a specific proposal: Analog Black Hole Evaporation via Lasers (AnaBHEL) — to use a laser-driven wakefield as a laboratory for black hole physics.

\subsection{Extreme Accelerators: The Plasma Wakefield and the Origin of Cosmic Rays}

\begin{figure}[h!]
    \centering
    \includegraphics[width=0.7\linewidth]{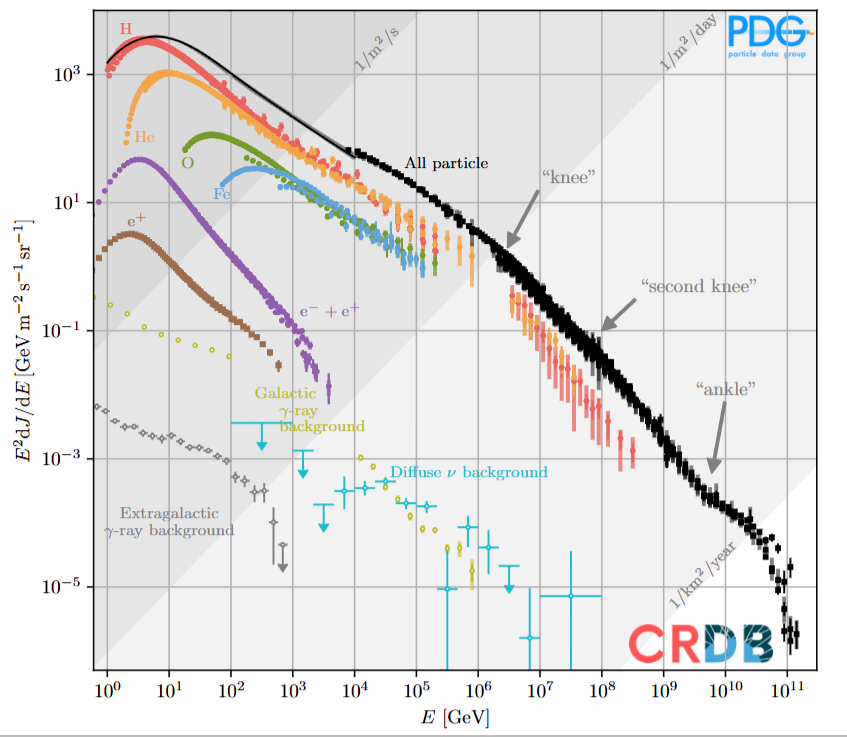}
    \caption{The all-particle cosmic ray energy spectrum, showing its characteristic power-law nature with features known as the "knee" and "ankle". The origin of particles beyond the ankle, the UHECRs, remains a major open question in astrophysics. Adapted from
  Ref.~\cite{PDG2024CosmicRays}, Fig.~30.1. © Particle Data Group 2024, CC-BY 4.0.}
    \label{fig:cosmic_ray_spectrum}
\end{figure}

One of the most enduring mysteries in astrophysics is the origin of ultra-high-energy cosmic rays (UHECRs), particles—mostly protons and atomic nuclei—that strike Earth's atmosphere with macroscopic energies exceeding $10^{18}$ eV, and in some cases approaching $10^{20}$ eV (see Fig.~\ref{fig:cosmic_ray_spectrum}). For over half a century, the leading theoretical paradigm for their acceleration has been the \emph{Fermi acceleration} mechanism \cite{fermi1949}.

In its modern form, known as diffusive shock acceleration (DSA), or the first-order Fermi acceleration, particles are thought to be accelerated by repeatedly crossing the shock front of a supernova remnant or an astrophysical jet. At each crossing, a particle gains a small amount of energy, proportional to its current energy ($\Delta E \propto E$). This stochastic process naturally leads to a power-law energy spectrum, $dN/dE \propto E^{-p}$, where the spectral index $p$ is predicted to be close to 2 for strong shocks. While remarkably successful in explaining the origin of cosmic rays up to the "knee" of the spectrum at $\sim 10^{15.5}$ eV, the Fermi mechanism faces significant challenges at the highest energies. To accelerate a particle to $10^{20}$ eV, the mechanism requires a vast accelerator size and a strong magnetic field to confine the particle long enough for gradual acceleration to occur. The maximum achievable energy is constrained by the famous "Hillas criterion" \cite{hillas1984}, and for UHECRs, only the most extreme astrophysical objects, such as Active Galactic Nuclei (AGN) or Gamma-Ray Bursts (GRBs), are plausible candidates. Even then, the acceleration process is relatively slow and may be inefficient at the highest energies.

The plasma wakefield accelerator offers a compelling alternative paradigm that can potentially overcome the limitations of the slow, gradual Fermi mechanism. In 2002, Chen, Tajima, and Takahashi proposed that the same PWFA physics studied in the laboratory could be the engine driving cosmic accelerators \cite{Pisin2002UHCR}. Their model posits that in the relativistic outflows and jets associated with AGNs or GRBs, large-scale plasma wakefields can be excited by Alfv\'en shocks or magnetosonic shocks.

These propagating medium waves or shocks, moving through the ambient plasma of the jet, can act as the "driver" of a wakefield, in a manner analogous to a laser or particle beam in the lab. The turbulent, magnetized plasma of the jet provides the medium for a powerful plasma wakefield to be established behind the front of the medium wave or shocks. Charged particles trapped in this wake can then be accelerated with extreme efficiency and over very short distances. A key insight of this mechanism is that the stochastic nature of particles encountering both accelerating and decelerating phases of these turbulent wakes naturally reproduces the observed power-law energy spectrum, with a predicted spectral index of $p=2$, consistent with observations.

The theoretical foundation of this "media-driven" wakefield acceleration has been robustly supported by particle-in-cell (PIC) simulations. Hoshino \cite{hoshino2008wakefield} and Chang et al.\cite{Chang2009Magetowave} demonstrated that relativistic, perpendicular magnetosonic shocks in electron-positron plasmas can efficiently generate nonlinear plasma waves and accelerate particles, confirming the core mechanism. Subsequent laboratory experiments have provided direct validation. By firing a high-power laser onto a magnetized plasma target, Kuramitsu et al. successfully observed the generation of a large-amplitude magnetosonic wave and the subsequent acceleration of electrons, a direct tabletop analog of the proposed astrophysical process \cite{Kuramitsu2012}.

This PWFA-based model for UHECRs not only offers a viable acceleration mechanism but also suggests potential observational signatures. For example, the acceleration process would be accompanied by strong synchrotron-like radiation from the accelerated particles, which could contribute to the observed non-thermal photon spectra from these astrophysical sources. The ongoing efforts to correlate the arrival directions of UHECRs with potential sources like AGNs, combined with multi-messenger observations in photons and neutrinos, will be crucial in testing this compelling new paradigm for nature's most powerful accelerators.

\subsection{Extreme Observers: Toward a Laboratory Test of the Unruh Effect}

Beyond the realm of cosmic accelerators, the combination of intense lasers and plasmas offers a tantalizing prospect: the ability to probe the very nature of spacetime and the vacuum itself. One of the most profound and counter-intuitive predictions arising from the synthesis of quantum field theory and relativity is the \emph{Unruh effect} \cite{Unruh1976}.

The effect states that an observer undergoing constant proper acceleration $a$ through what an inertial observer would perceive as empty Minkowski spacetime (a vacuum) will instead perceived by the accelerating observer as being immersed in a thermal bath of particles. The vacuum, from the perspective of this ``extreme observer," appears to have a finite temperature, known as the Unruh temperature:
\begin{equation}
    k_B T_U = \frac{\hbar a}{2\pi c},
    \label{eq:unruh_temp}
\end{equation}
where $k_B$ is the Boltzmann constant and $\hbar$ is the reduced Planck constant. This prediction is deeply connected to the celebrated Hawking radiation from black holes \cite{Hawking1974}. Through the Einstein equivalence principle, the physics experienced by a uniformly accelerating observer is locally indistinguishable from that of a stationary observer in a uniform gravitational field. The Unruh effect is thus the flat-spacetime analog of Hawking radiation, and its experimental verification would provide profound confirmation of our understanding of quantum fields in non-inertial frames, a cornerstone of quantum gravity theories.

The physical origin of the Unruh effect lies in the observer-dependent nature of the particle concept in quantum field theory. An inertial observer decomposes the quantum field into positive and negative frequency modes, defining the vacuum as the state with no positive-frequency particles. However, the trajectory of a uniformly accelerating observer defines a different set of natural \emph{Rindler} modes. The inertial vacuum state, when expressed in this new basis of Rindler modes, is no longer empty but is found to be a thermal state, corresponding to a blackbody spectrum at the Unruh temperature.

Despite its fundamental importance, the Unruh effect has never been directly observed. The primary reason is the immense acceleration required to produce a measurable temperature. For an acceleration of $a \approx 10^{20} \text{ m/s}^2$, the Unruh temperature is only about $0.4$ Kelvin. Generating and sustaining such accelerations for a macroscopic detector is technologically impossible.

An alternative approach is to treat a high-energy electron itself as the \emph{accelerated observer} and to search for signatures of its interaction with the Unruh thermal bath.
One key insight, proposed by Bell and Leinaas ~\cite{Bell1983}, is to reinterpret the well-known 
Sokolov-Ternov effect in an electron-sychrotron, where the spin-flip polarizations are not equal between spin-up and spin-down at equilibrium, as a signature of the Unruh effect. Specifically, the thermal photons perceived by the accelerating electron alter the Sokolov-Ternov transition probability, so that the equilibrium longitudinal polarization acquires a tiny correction $\delta P_L / P_L \sim \mathcal{O}(T_U/m_ec^2)$.

\begin{figure}[h!]
    \centering
    \includegraphics[width=0.5\linewidth]{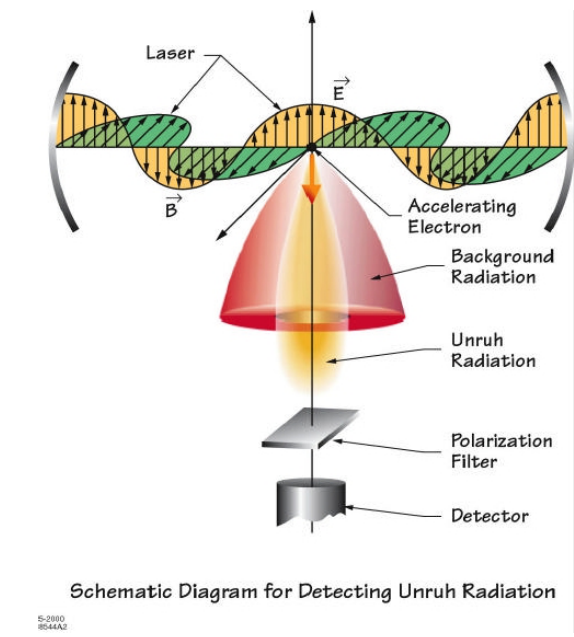}
    \caption{A schematic concept for detecting Unruh radiation. A test electron is trapped and accelerated by the intense electric field at an antinode of a laser standing wave inside a cavity. The Unruh temperature experienced by the electron would cause a minute, but in principle detectable, modification to its spin dynamics in an external magnetic field.}
    \label{fig:schematic_unruh_exp}
\end{figure}

Chen and Tajima took a different route. They proposed an experimental realization of the Unruh effect using the intense electric field of a standing wave formed by two counter-propagating ultra-intense lasers inside a cavity for an extremely intense linear acceleration of a test electron, the observer ~\cite{Chen1999Unruh}. Instead of relying on the internal degree of freedom, i.e., the spin, of the observer, Chen and Tajima invoke the spacetime degree of freedom where the stochastic fluctuations of the accelerating electron as a result of its interaction with the thermal heat bath would trigger a ``Unruh radiation". As sketched in Fig.~\ref{fig:schematic_unruh_exp}, the electron is trapped near a field antinode where it experiences, for example, an acceleration $a\gtrsim10^{21}\ \mathrm{m\,s^{-2}}$, which gives rise to a Unruh temperature of $T_U\sim4\ \mathrm{K}^{\circ}$ for the thermal spectrum of Unruh radiation.

While conceptually elegant, this approach faces its own formidable experimental challenges, including the difficulty of isolating the extremely subtle Unruh signal from overwhelming classical background radiation (Larmor radiation) and other noise sources. This difficulty has motivated the search for alternative experimental platforms where the signatures of such general relativistic effects might be more pronounced and accessible. This leads us directly to the concept of analog gravity and the AnaBHEL proposal, which aims not to accelerate a single particle detector, but to accelerate a macroscopic plasma mirror to simulate the event horizon itself, a topic we will explore in detail in the next chapter.

\section{Wakefield as a Laboratory for Black Hole Physics: The AnaBHEL Proposal}
\label{sec:AnaBHEL}

\subsection{The Information Loss Paradox and the Rise of Analog Gravity}

The preceding sections have traced the journey of plasma wakefield acceleration from its theoretical inception to its realization in state-of-the-art experimental facilities. We now arrive at the ultimate frontier, where this powerful tool for accelerating particles transforms into a laboratory for probing the deepest questions in fundamental physics. The central motivation for this leap is one of the most profound theoretical conflicts of the modern era: the black hole information loss paradox \cite{hawking1976, Polchinski2017}.

The paradox emerges from the collision of two pillars of modern physics. According to quantum mechanics, the evolution of a closed system must be unitary, meaning the information is never lost. A system that begins in a pure quantum state (like a star described by a single wavefunction) must evolve into another pure state. However, Stephen Hawking's celebrated 1974 calculation, which combined general relativity with quantum field theory, showed that black holes are not truly black but radiate thermally, a phenomenon now known as Hawking radiation \cite{Hawking1974}. This thermal radiation is described by a mixed state, a statistical ensemble that lacks complete information of the initial state. As the black hole evaporates completely, it seems to transform the initial pure state into a final mixed state of thermal radiation, constituting a violation of unitarity. This conflict strikes at the very heart of our understanding of Nature, and its resolution is widely believed to require a full theory of quantum gravity.

Unfortunately, direct observation of Hawking radiation from astrophysical black holes is far beyond current technologies. The Hawking temperature of a solar-mass black hole is orders of magnitude below the cosmic microwave background, and its evaporation timescale is much longer than the age of the universe. This observational impasse has given rise to the vibrant field of \emph{analog gravity}, pioneered by Unruh in 1981 \cite{unruh1981}. The core idea is that phenomena like Hawking radiation can be simulated in accessible laboratory systems where perturbations (such as sound waves or light) propagate in a moving medium. If the equations governing these perturbations are mathematically analogous to those for quantum fields near a black hole, then the laboratory system can serve as a semi-classical quantum gravity simulator.

Over the past two decades, this field has flourished, with remarkable experiments observing Hawking-like phenomena in various systems, including sound waves in Bose-Einstein condensates and water tanks, light in optical fibers, and excitations in superconducting circuits \cite{nation2009,lahav2010,steinhauer2014,steinhauer2019}. These experiments have brilliantly confirmed the robustness of Hawking's semi-classical calculations in analogous systems.

The Analog Black Hole Evaporation via Lasers (AnaBHEL) proposal, introduced by Chen and Mourou in 2017 \cite{pisin2017AnaBHEL}, enters this landscape with a unique and powerful approach rooted in the physics of relativistic plasma wakefields \cite{Bulanov2013review}. Unlike fluid or condensed matter systems where the analog radiation consists of phonons or other quasi-particles, the AnaBHEL experiment aims to simulate the production of actual photons from the vacuum, triggered by a boundary, a \emph{flying plasma mirror} accelerating at relativistic speeds. This offers a potentially cleaner and more direct analog to the fundamental quantum field theory processes near a real event horizon. 

\begin{figure}[h!]
    \centering
    \includegraphics[width=1.0\linewidth]{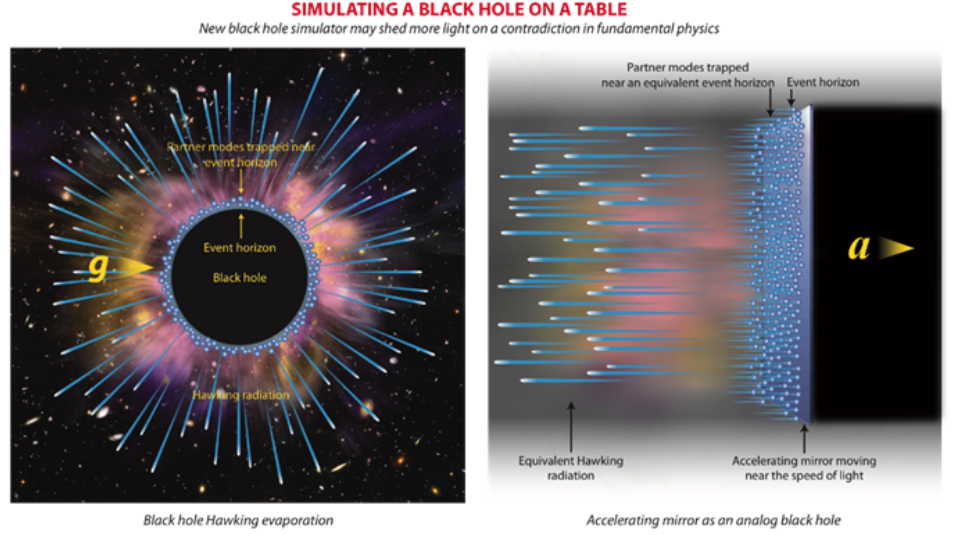}
    \caption{Conceptual analogy between black hole evaporation and the accelerating mirror model. \textbf{(Left)} A black hole's immense gravity ($g$) creates an event horizon, from which Hawking radiation is emitted while entangled partner modes are trapped. \textbf{(Right)} An accelerating mirror creates an analogous "event horizon" in its own reference frame, producing a thermal spectrum of photons (equivalent Hawking radiation) from the vacuum. Einstein's {\bf equivalence principle} provides an intuitive link between the gravitational field ($g$) and the mirror's acceleration ($a$).}
    \label{fig:equivalent_principle}
\end{figure}

The connection between these two disparate systems can be intuitively understood through Einstein's equivalence principle: the physics in a uniformly accelerating reference frame is indistinguishable from that in a uniform gravitational field. Plasma wakefields are renowned for sustaining some of the most extreme accelerations ever produced in a laboratory. This capability, therefore, provides an unique platform to simulate phenomena associated with extreme gravity, such as black hole event horizons, in a controlled terrestrial setting, as conceptually illustrated in Fig.~\ref{fig:equivalent_principle}. The following sections will detail the theoretical foundations, experimental design, and scientific goals of this ambitious proposal.

\subsection{Relativistic Flying Mirrors}

At the heart of the AnaBHEL proposal lies a powerful yet conceptually simple theoretical framework within analog gravity: the moving mirror model. This model is a specific manifestation of a more general phenomenon known as the Dynamical Casimir Effect (DCE), in which non-adiabatic changes to a quantum system's boundary conditions or environment can convert virtual vacuum fluctuations into real, observable particles \cite{Dodonov2020Physics}. The idea that an accelerating boundary could create particles from the vacuum has a long history, with foundational work on the moving mirror model by Moore, Davies, and Fulling in the 1970s \cite{Moore1970, FullingDavies1976}. They demonstrated that the non-uniform acceleration of the mirror perturbs the vacuum state, leading to the creation of real particles.

This process arises from the mixing of positive and negative frequency modes of the quantum field upon reflection from the moving boundary, a phenomenon mathematically described by a Bogoliubov transformation. The crucial insight is that the spectrum of the created particles is intimately tied to the mirror's trajectory. Remarkably, specific trajectories were found to produce a perfectly thermal, Planckian spectrum of particles, directly mimicking the thermal nature of Hawking radiation \cite{DaviesFulling1977, CarlitzWilley1987PRD}. Later work, for instance by Wilczek, further explored this connection, showing how different trajectories could be engineered to model various aspects of the evaporation process, including the back-reaction on the mirror's trajectory from the emitted radiation \cite{Wilczek1993}. The moving mirror model thus became a cornerstone of analog gravity, providing a clean theoretical laboratory to study particle creation in dynamic spacetime.

While the relativistic moving mirror remained a theoretical ideal, the broader field of DCE saw landmark experimental success. Using superconducting circuits and Josephson metamaterials, which allow for the rapid modulation of a microwave cavity's effective electrical length, several groups have unambiguously observed the creation of microwave photon pairs from the vacuum, confirming the core prediction of the DCE \cite{Wilson2011Nature, Lahteenmaki2013PNAS}. These experiments established an operational bridge between the theory of quantum radiation from moving boundaries and realizable laboratory platforms. However, these systems operate in the non-relativistic regime. The central challenge of creating a boundary moving with the extreme relativistic accelerations required to produce a detectable thermal flux analogous to Hawking radiation remained.

The answer to this challenge emerged from the field of high-intensity laser-plasma interactions. As we have seen, when an ultra-intense laser pulse propagates through an underdense plasma in the blowout regime, it creates a wake structure where electrons are violently expelled and then rush back to form a thin, extremely dense shell at the rear of the first bubble \cite{Pukhov2002}. Pioneering theoretical work by Bulanov, Esirkepov, and collaborators proposed that this dense, co-propagating electron shell could act as a relativistic, reflective boundary—a flying plasma mirror (FPM) \cite{Bulanov2003}. The key properties of an FPM make it a near-ideal candidate for realizing the moving mirror model:
\begin{itemize}
    \item \textbf{Relativistic Velocity}: The FPM is driven by the laser pulse and co-propagates at a velocity close to the laser's group velocity in the plasma, which can be highly relativistic ($\gamma \gg 1$).
    \item \textbf{High Reflectivity:} The electron density in the shell can be orders of magnitude higher than the background plasma density, potentially exceeding the critical density for an incident probe laser pulse. This allows the shell to act as a highly reflective mirror, especially for lower-frequency light.
    \item \textbf{Controllable Trajectory:} As will be detailed in the next section, the mirror's velocity is fundamentally linked to the plasma density. This provides a direct experimental knob to control the mirror's trajectory, inducing the acceleration necessary for particle creation.
\end{itemize}

This theoretical concept was soon followed by compelling experimental validation. A series of landmark experiments led by M. Kando and collaborators provided the first direct proof of FPMs. They successfully reflected a probe laser pulse off these laser-generated plasma structures, observing a dramatic frequency up-shift of the reflected light. The magnitude of this up-shift, resulting from a double-Doppler effect at the relativistic mirror, was consistent with theoretical predictions and confirmed the existence of these relativistic, reflective boundaries in the laboratory \cite{Kando2007, Kando2009}.

The confluence of the well-established moving mirror theory with the experimental discovery of the flying plasma mirror is the key ingredient of the AnaBHEL proposal. However, flying plasma mirrors moving with uniform velocity do not mimic the Hawking effect. Only those FPMs with nontrivial trajectories would be able to simulate it. The insight of Chen and Mourou is to vary the plasma density so as to tune the refractive index and thus the laser group velocity.  It provides a concrete, physically realizable path to engineer an accelerating boundary in the laboratory and, for the first time, to directly test the predictions of particle creation from a controlled, relativistic trajectory.

\subsection{From Plasma Density to a Thermal Spectrum: The Theoretical Link}

The discovery of the flying plasma mirror provided the physical boundary in the moving mirror model, but the crucial element for simulating Hawking radiation is not just motion, but a specific form of acceleration. The theoretical work that makes AnaBHEL a quantitative and experimentally testable proposal is the establishment of a direct, deterministic link between the FPM's trajectory and the background plasma density profile it traverses. This connection provides the experimentalist with a "knob" to precisely dial in the desired relativistic motion.

In a uniform plasma, a stable FPM travels at a near-constant velocity, close to the laser's group velocity. However, the mirror's velocity, $v_M$, is fundamentally tied to the evolution of the wake structure, which in turn depends on the local plasma wavelength, $\lambda_p \propto 1/\sqrt{n_e}$. As the FPM propagates through a plasma with a longitudinal density gradient ($dn_e/dx \neq 0$), the wake structure evolves, causing the mirror to accelerate or decelerate. While several factors can influence the mirror's motion (e.g., laser energy depletion), theoretical analysis shows that for typical parameters, the plasma density gradient is the dominant control mechanism \cite{pisin2020mirrorTrajectory}.

The theoretical development was the identification of a specific plasma density profile that would guide the FPM onto a trajectory known to produce a perfectly thermal emission spectrum. Theoretical work by Chen and Mourou established that if the background plasma electron density, $n_e(x)$, is engineered to follow a specific ``one-plus-exponential" form,
\begin{equation}
    n_e(x) = n_{e0}\left(1 + b e^{-x/D}\right)^2,
    \label{eq:anabhel_density}
\end{equation}
where $n_{e0}$ is the background density, $b$ is the amplitude of the density bump, and $D$ is the characteristic scale length of the exponential decay, then the resulting trajectory of the FPM, $x_M(t)$, in the asymptotic late-time limit, takes the form:
\begin{equation}
    x_M(t) \approx ct - A e^{-ct/D} + B,
    \label{eq:anabhel_trajectory}
\end{equation}
where A and B are constants. This trajectory is mathematically identical to the Davies-Fulling (DF) trajectory, which is known from first-principle quantum field theory calculations to produce a perfectly thermal spectrum of particles from the vacuum \cite{DaviesFulling1977}.

This remarkable connection directly forges a link between a controllable experimental parameter, the characteristic scale length $D$ of the plasma density gradient, and the effective temperature of the emitted analog Hawking radiation, $T_H$. The relationship is derived by identifying the effective surface gravity of the analog system as $\kappa = c/(2D)$. This yields an analog Hawking temperature:
\begin{equation}
    k_B T_H = \frac{\hbar \kappa}{2\pi} = \frac{\hbar c}{4\pi D}.
    \label{eq:anabhel_temp}
\end{equation}
This equation represents the theoretical heart of AnaBHEL. It is a powerful \emph{dictionary} that translates the language of experimental plasma engineering into the language of fundamental quantum gravity. It transforms the abstract challenge of simulating Hawking radiation into the concrete, albeit demanding, experimental task of fabricating and characterizing a gas target with a precisely shaped density gradient on a sub-micrometer scale. For example, to produce a detectable thermal spectrum in the infrared range (corresponding to a peak wavelength of $\lambda_{\text{peak}} \sim 10\,\mu\text{m}$), the required scale length is $D \approx 0.5\,\mu\text{m}$. This stringent requirement is the primary driver behind the gas target R\&D efforts that are central to the experimental realization of AnaBHEL.

\subsection{Experimental Realization and Quantum Signatures}

The theoretical framework linking plasma density to a thermal spectrum provides a clear blueprint for an experiment. The AnaBHEL proposal translates this blueprint into a tangible experimental design, outlining a path towards not only the first observation of Hawking-like radiation from a flying mirror but also a direct probe of the quantum information paradox.

\begin{figure}[!ht]
    \centering
    \includegraphics[width=0.9\linewidth]{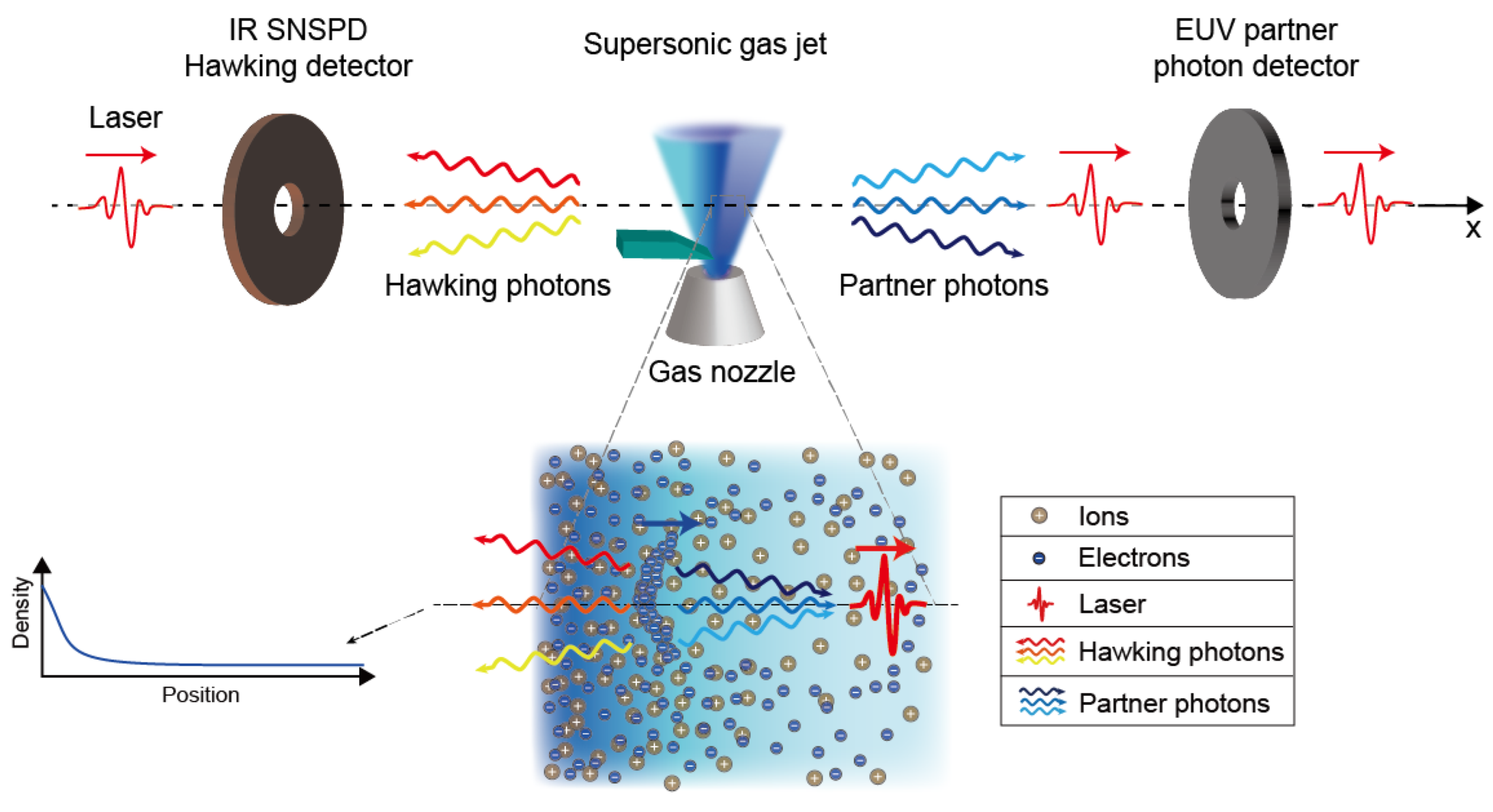} 
    \caption{Conceptual schematic of the AnaBHEL experiment. An intense laser pulse interacts with a structured gas jet target, creating a sharp density gradient via a shock wave. This generates a relativistic flying plasma mirror (FPM). Analog Hawking photons emitted backward are red-shifted (e.g., to IR) and detected. Forward-propagating partner photons retain higher energy (e.g., EUV) and are detected separately. The quantum correlation between these photon pairs is the key observable for investigating the information loss paradox. (Adapted from Ref.~\cite{pisin2022anabhel}).}
    \label{fig:anabhel_schematic}
\end{figure}

The proposed setup for AnaBHEL, shown schematically in Fig.~\ref{fig:anabhel_schematic}, is designed to precisely engineer the laser-plasma interaction required to generate the accelerating flying plasma mirror (FPM). An ultra-intense driving laser pulse (typically with petawatt-class peak power) is focused onto a highly structured gaseous target. As established in the previous section, the central technological challenge is the creation of the "one-plus-exponential" plasma density profile upon ionization. The most promising method for producing the required sharp, sub-micrometer density gradient is to utilize the shock front generated by the interaction of a supersonic gas jet with a sharp obstacle, such as a thin wire or a knife-edge blade. Extensive research involving computational fluid dynamics (CFD) simulations and high-resolution optical diagnostics has been dedicated to designing and characterizing such targets \cite{YK2024}, as this is the critical enabling technology for the entire experiment.

As the laser-driven FPM forms and propagates down this density gradient, its prescribed acceleration forces the emission of particle pairs from the quantum vacuum.

The primary scientific objective of AnaBHEL is the simultaneous detection of the particle pairs created by the accelerating mirror. These two particles, the ``Hawking photon" and its ``partner photon," are predicted to have vastly different observational signatures due to the relativistic kinematics of their emission from the near-luminal mirror.

\paragraph{Analog Hawking Photons} The photons emitted backward relative to the FPM's motion are the direct analog of the Hawking radiation that escapes to infinity from a black hole. Due to the relativistic Doppler shift from the receding source, their frequency is significantly downshifted. For the parameters required to achieve a detectable temperature ($D \sim 0.5\,\mu\text{m}$), these analog Hawking photons are predicted to lie in the infrared (IR) spectrum (e.g., $\lambda \sim 10\,\mu\text{m}$). Their detection requires highly sensitive, low-noise detectors capable of registering single photons, such as a superconducting nanowire single-photon detector (SNSPD) \cite{HsinYe2025SNSPD}.

\paragraph{Probing Unitarity with Partner Photons}
The ultimate and most profound goal of AnaBHEL extends beyond simply observing a thermal spectrum. The information loss paradox is fundamentally a question about quantum information and entanglement. According to theory, for every Hawking photon that escapes, an entangled partner particle is also created. In the black hole case, this particle falls across the horizon, carrying away the "negative energy" required for mass loss. In the moving mirror analog, the partner photon is predicted to propagate forward. In the semi-transparent FPM model, this partner can pass through the mirror, largely unaffected by the relativistic kinematics that so drastically redshift the backward-propagating Hawking photon \cite{pisin2022anabhel}. These partner photons are therefore predicted to be in the extreme ultraviolet (EUV) or even soft X-ray range, requiring a separate suite of diagnostics for their detection.

Because analog gravity experiments are realized in laboratory systems governed by standard quantum field theory, the global evolution of the entire system must be unitary. The profound scientific question is not "whether" information is conserved, but "how" it is encoded in the correlations between the outgoing radiation and its partners \cite{pisin2022anabhel}. The moving mirror model makes this question experimentally accessible. It has been shown theoretically that different mirror trajectories, $z(t)$, generate distinct entanglement entropy histories for the Hawking radiation sector. While an eternal thermal trajectory (e.g., Carlitz-Willey) leads to a monotonically increasing entanglement entropy, other trajectories that model a fully evaporating, horizonless object can reproduce the celebrated "Page curve," where the entanglement entropy first rises and then falls, consistent with a pure final state \cite{CarlitzWilley1987PRD, GoodJHEP2017, AkalPRL2021, ChenYeom2017}.

AnaBHEL is designed to probe this very physics. By detecting the IR Hawking photons and the EUV partner photons in coincidence on a shot-by-shot basis, the experiment aims to measure the quantum correlations between them. The observation of entanglement would provide direct experimental evidence that the complete final state (Hawking + partner photons) remains pure, even though each subsystem alone appears thermal. This would be a landmark result, offering powerful experimental support for the principle of unitarity in the context of black hole evaporation.

It is this unique capability to probe the quantum information content of the radiation that distinguishes AnaBHEL as a second-generation analog gravity experiment. It is designed not just to see the thermal glow of Hawking radiation, but to read the information encoded within its quantum correlations. It is important to note, however, that not all high-energy radiation from the mirror corresponds to the Hawking partners. As shown by Hotta et al., a final burst of radiation from a suddenly stopping mirror is a classical shock effect and should not be confused with the quantum partner modes, providing a clear criterion for experimental design and interpretation \cite{HottaPRD2015, OsawaPRD}.

\subsection{Challenges and Perspective: A New Frontier for Wakefield Physics}

Realizing the ambitious scientific goals of AnaBHEL is an attempt that pushes the very limits of current technology in high-intensity lasers, plasma target R\&D, and single-photon detection. Overcoming these challenges is the central focus of the ongoing collaborative effort. One of the primary and demanding challenges is the fabrication and characterization of the gas target. As established by Eq.~\eqref{eq:anabhel_temp}, the ability to generate a detectable Hawking temperature hinges on creating a stable, precisely shaped plasma density profile with a sub-micrometer scale length. Any deviation from the ideal "one-plus-exponential" form can alter the mirror's trajectory, distort the thermal spectrum, and compromise the interpretation of the results. This places extreme demands on the design of supersonic nozzles and shock-generation mechanisms, as well as on the high-resolution diagnostics required to validate them.

Furthermore, the theoretical model of a "perfectly reflecting" mirror is an idealization. A realistic flying plasma mirror is a semi-transparent object of finite thickness and transverse size. Recent theoretical work has begun to explore these crucial real-world effects. The partial transmission of the mirror, its finite size, and the frequency dependence of its reflectivity will inevitably modify the emitted spectrum, causing it to deviate from a perfect Planckian distribution \cite{KuanNan2024}. Quantifying the FPM's reflectivity, which depends on the peak electron density and thickness of the electron shell, is therefore critical. Estimates based on realistic plasma parameters through PIC simulation suggest a low reflectivity, potentially in the range of $R \sim 10^{-3} - 10^{-5}$ \cite{liu2021reflectivity}, which directly impacts the expected photon yield.

\begin{figure}[h!]
    \centering
    \includegraphics[width=0.7\linewidth]{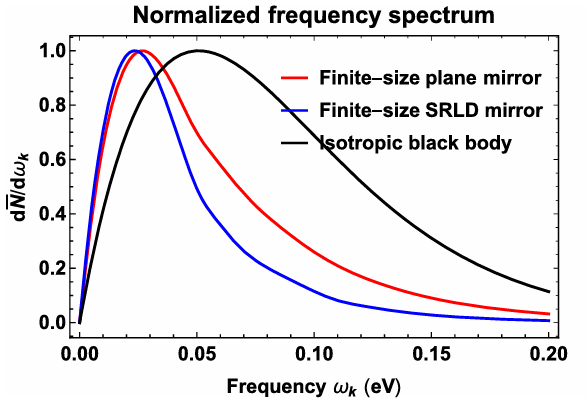}
    \caption{Normalized particle emission spectrum $dN/d\omega_k$ versus frequency $\omega_k$ for different mirror models. The black curve shows an ideal isotropic blackbody spectrum. The red curve represents emission from a finite-transverse-size, perfectly reflecting mirror following a DF trajectory. The blue curve incorporates both finite size and finite thickness effects using the square-root Lorentzian distribution (SRLD) model for the electron shell, which includes frequency-dependent reflectivity \cite{liu2021reflectivity}. The distortion from the ideal blackbody shape, particularly the shift towards lower frequencies, highlights the importance of realistic modeling. Adapted from \cite{KuanNan2024}}
	\label{fig:thermal_spectrum_kuanNan}
\end{figure}

This low reflectivity contributes to another major hurdle: the extremely low expected signal. Realistic estimates predict a yield of only $\sim 0.3$ analog Hawking photons per petawatt-class laser shot \cite{KuanNan2024}. Detecting such a faint signal requires extended experimental campaigns coupled with state-of-the-art, highly efficient single-photon detectors in the desired band. Moreover, this faint signal must be meticulously discriminated from the intense and complex background noise—from both classical Larmor radiation and plasma self-emission, which are inherent to the laser-plasma interaction environment. This necessitates sophisticated shielding, spectral filtering, and coincidence detection schemes.

Despite these formidable challenges, the scientific payoff is extraordinary, and significant progress is being made across all fronts, from target development to detector R\&D, as documented by the AnaBHEL collaboration \cite{pisin2022anabhel}. The project represents a philosophical culmination of the intellectual journey through wakefield physics. It elevates the plasma wakefield from a powerful tool for particle acceleration to a physical system in its own right—a controllable, relativistic spacetime analog. This new frontier allows us to ask profound questions about the quantum nature of our universe in a controlled laboratory setting. The ongoing efforts to realize AnaBHEL are therefore not just about building a single experiment; they are about opening an unprecedented window onto the quantum nature of gravity.

\section{Expanding Horizons and Future Outlook}
\label{sec:outlook}

The journey from the foundational principles of plasma wakefield acceleration to the ambitious proposal of AnaBHEL showcases a remarkable intellectual arc—one that elevates a novel accelerator concept into probing the quantum nature of gravity. Having explored this profound frontier, we now broaden our perspective to survey the other exciting horizons that the physics of plasma wakefields is opening up. The same principles of generating immense collective fields in a medium, which empower the AnaBHEL concept, also drive the quest for next-generation particle colliders, novel radiation sources, and even more exotic acceleration schemes. This final section will explore these expanding horizons, from the push towards ever-higher frequencies in solid-state systems to the practical roadmaps for future applications, culminating in a perspective on the future of this vibrant field.

\subsection{Towards Higher Fields: Wakefields in Crystals and Dielectrics}

The accelerating gradient of a plasma wakefield, $E_{\text{WB}} \propto \sqrt{n_e}$, points to a clear path for achieving even higher gradients: moving from gaseous plasmas to media with much higher electron densities. This progression leads to the exploration of solid-state systems, which represent both a near-term, practical alternative and a far-future, ultimate-gradient frontier.

\subsubsection{Dielectric Wakefield Accelerators (DWAs)}

A highly promising and experimentally mature approach is the use of dielectric wakefield accelerators (DWAs). In a DWA, a particle beam travels through a hollow channel in a dielectric material, such as fused silica or diamond \cite{Tremaine1997}. The beam's field polarizes the material, inducing a wakefield that can accelerate a trailing bunch. While the fields are generally lower than in a plasma, DWAs offer compelling advantages: they are passive, solid-state structures that can be manufactured with lithographic precision, offering the potential for excellent stability and control.

Significant experimental progress has established DWAs as a viable technology. Gradients of $\sim 1.3\,\text{GV/m}$ have been sustained over 15-cm structures, and breakdown thresholds of the inner dielectric walls have been measured to be as high as $\sim 14\,\text{GV/m}$, defining realistic operating windows \cite{OShea2016, Thompson2008}. Crucially, DWAs have served as a key platform for demonstrating the principles of high-efficiency energy transfer. Using a ramped bunch train at the Argonne Wakefield Accelerator (AWA), a transformer ratio of $R \approx 3.4$ was achieved \cite{Jing2007PRL, Jing2011PRSTAB}. Even more impressively, a recent experiment at AWA used an emittance-exchange beamline to shape a single drive bunch into a quasi-triangular profile. This shaped bunch directly produced a measured transformer ratio of $R \gtrsim 4.5$, providing a landmark validation that single-bunch current-profile shaping is a practical route to overcoming the $R \le 2$ limit in structure-based accelerators \cite{Gao2018PRL}.

The central challenge for DWAs remains beam stability. Strong transverse wakefields can drive beam-breakup instabilities in the driver and cause emittance growth in the witness bunch. An active area of research is the design of advanced structures (e.g., planar or two-channel geometries) and the use of elliptical beams to mitigate these transverse effects \cite{Saveliev2022PRAB}. Beyond acceleration, DWA structures are also finding practical use as passive "dechirpers" to control the longitudinal phase space of electron beams, including those generated by LWFAs \cite{Nie2018NIMA}. DWAs thus represent a fascinating and rapidly maturing bridge between conventional accelerator structures and plasma-based schemes.

\subsubsection{Crystal and Nanostructure Acceleration}

At the ultimate gradient frontier lies the concept of a solid-state plasma accelerator, a vision of harnessing the immense electron densities of solids to produce unprecedented accelerating gradients. Interestingly, the historical duality of drivers that defines the field of plasma acceleration, particle bunches versus laser pulses, was found at its very inception in this new domain. In the mid-1980s, two visionary proposals emerged concurrently: Chen and Noble proposed a crystal wakefield accelerator driven by a relativistic particle beam, a direct analogue to the PWFA \cite{chen1986solid, Chen1987Crystal}, while almost simultaneously, Tajima and Cavenago proposed a crystal X-ray accelerator, suggesting an X-ray laser could serve as the driver, in the spirit of the LWFA \cite{Tajima1987Crystal}. Both concepts are built on the same compelling foundation: by leveraging the valence electron density of a solid ($n_e \sim 10^{23}\,\text{cm}^{-3}$), these schemes promise gradients on the scale of TV/cm. Furthermore, the regular lattice structure of a crystal could provide a "channeling" effect, guiding accelerated particles along atomic planes while mitigating emittance growth from multiple scattering. This effect is particularly pronounced for positively charged particles, which experience a strong focusing field from the lattice ions.

For decades, this concept remained a far-future vision due to technological hurdles, primarily the rapid material damage from the intense driver and the strong radiation damping experienced by the channeled particles. However, subsequent theoretical work revealed that this strong damping, typically a detriment, could be an advantage in a crystal. In a continuous focusing channel, the particle's transverse motion is cooled by radiation emission, a phenomenon known as \emph{absolute damping}~\cite{HuangChenRuth1995}. Following their formalism, the transverse action $J$ of a particle in a channel with restoring constant $K$ decays exponentially:
\begin{equation}
\frac{dJ}{dt} = -\frac{2 r_e K}{3 m_e c} J,
\label{eq:abs_damp}
\end{equation}
where $r_e$ is the classical electron radius. Crucially, because the transverse recoil from photon emission is absorbed by the crystal lattice as a whole (analogous to the M\"ossbauer effect), there is no associated random quantum excitation. The particle inevitably cools to its transverse ground state, achieving a minimum normalized emittance limited only by the uncertainty principle:
\begin{equation}
\epsilon_{n, \text{min}} = \frac{\hbar}{2 m_e c} = \frac{\bar{\lambda}_{_C}}{2},
\label{eq:min_emmitance}
\end{equation}
where $\bar{\lambda}_{_C}$ is the reduced Compton wavelength. The prospect of producing such quantum-limited emittance beams, orders of magnitude smaller than what is achievable with conventional methods, offers a path toward the extremely high luminosities required for future particle colliders and is a key driver for the renewed interest in this research area.

This renewed interest is fueled by recent advances in high-brightness X-ray sources (like XFELs), the fabrication of advanced nanostructures such as carbon nanotubes, and sophisticated computational modeling. Modern theoretical studies, for instance, have explored X-ray wakefield acceleration and the accompanying betatron radiation within nanotubes, providing a more concrete picture of the underlying physics~\cite{Zhang2016Xray}. New experimental programs, such as FACET-II E336, are now revisiting the feasibility of channeling acceleration in this new context, aiming to explore the unique physics of solid-density plasmas and to quantify the fundamental limits imposed by radiation and material damage~\cite{Gilljohann2023JINST}. While still a long-term prospect, crystal and nanostructure acceleration continues to represent the conceptual endpoint in the quest for the highest possible accelerating gradients, pushing the boundaries of what is thought possible in accelerator science.

\subsection{Redefining the Figure of Merit: The Quantum Luminosity}
For any particle collider, the ultimate figure of merit is not just the center-of-mass energy, but the \emph{luminosity}, $\mathcal{L}$, which quantifies the rate of collisions and thus the potential for scientific discovery. Luminosity is classically defined by the geometry and density of the colliding bunches: 
\begin{equation}
\mathcal{L}_{CL}\equiv f_{rep}\frac{N^2}{4\pi\sigma_x\sigma_y},
\label{eq:classical_Luminosity}
\end{equation}
where $f_{rep}$ is the repetition rate of collisions, $N$ the number of particles and $\sigma_x$ and $\sigma_y$ are the transverse sizes of the colliding beams. In this mode of operation in conventional high energy colliders, the particles in the beam are randomly distributed. Therefore, the probability of collision of two particles from the colliding beams is low, somewhat similar to the situation of two colliding galaxies. We may refer to the conventional concept of luminosity the {\it classical luminosity}.

The absolute damping of high-energy particles to the ground state of the transverse momentum in a continuous focusing channel reviewed in the previous sub-section inspired Chen and Noble \cite{Chen1987Crystal} to introduce a new concept in high energy physics, the {\it quantum luminosity}. Let us envision that high energy particle beams are injected to a section of crystal from opposite sides before collision. The absolute damping effect would confine and reduce the transverse momenta of all particles to their ground state before they collide with the oncoming particles that propagate through the same crystal channels from the other end. The original random distribution of beam particles is now replaced by a honeycomb of $n_{ch}$ parallel channels, with $n_b$ particles propagating in each channel. The luminosity then becomes
\begin{equation}
\mathcal{L}_{QM}\equiv f_{rep}n_{ch} \frac{N_b^2}{4\pi\epsilon_{n, min}^2}=f_{rep}\cdot \frac{\pi r_c^2}{4\pi\sigma_x\sigma_y}\cdot \frac{N^2}{\pi \bar{\lambda}_{_C}^2}=\frac{r_c^2}{\bar{\lambda_c^2}}\mathcal{L}_{CL},
\label{eq:quantum_Luminosity}
\end{equation}
where $n_{ch}=(4\pi\sigma_x\sigma_y)/\pi r_c^2$
is the number of crystal channels covered by the jnjected beam spot size, $r_c$ is the channel size, and $N_b=N/n_{ch}$ is the number of particles in each channel. It is called quantum luminosity because it involves Planck's constant $\hbar$. We see that the ratio between $\mathcal{L}_{QM}$ and $\mathcal{L}_{CL}$ is equal to the ratio between the crystal channel size and Compton wavelength. Typical crystal channel size is of the order nanometer, while Compton wavelength $\bar{\lambda}_c=3.8\times 10^{-11}{\rm cm}$. Therefore, $r_c^2/\bar{\lambda}_{_C}^2\sim 10^6-10^8$, which would be a huge gain!

\subsection{Applications and the Global Roadmap}
The immense accelerating gradients and ultrashort bunch durations inherent to plasma wakefield accelerators are not just scientifically fascinating; they are enabling technologies that promise to revolutionize a vast range of scientific, medical, and industrial applications. By dramatically shrinking the size and cost of high-energy particle accelerators, PWFA and LWFA technologies are poised to democratize access to high-energy beams and spawn new capabilities. This vision is now being translated into concrete, strategic roadmaps by the international scientific community. 

\subsubsection{A New Generation of Compact, Brilliant Light Sources}

Perhaps the most immediate and tangible application of plasma accelerators is in driving a new generation of compact, brilliant radiation sources.
\begin{itemize}
    \item \textbf{Free-Electron Lasers (FELs):} Conventional X-ray FELs are kilometer-scale user facilities. Plasma accelerators offer a path to reduce their footprint to the scale of a large room. Landmark experiments have already demonstrated FEL lasing in the EUV and soft X-ray regimes driven by both LWFA \cite{Wang2021} and PWFA \cite{Pompili2022} beams. The ongoing challenge is to improve the stability, repetition rate, and beam quality to a level required for a true user facility, a primary goal for facilities like DESY's FLASHForward. These pioneering achievements have firmly established the viability of plasma-based FELs, transitioning them from conceptual possibilities to near-term realities. The current status, technological challenges, and promising prospects for developing user-ready, compact FELs powered by plasma accelerators are thoroughly examined in a recent review by Galletti et al. \cite{Galletti2024}.

    \item \textbf{Plasma-Based and Hybrid Radiation Sources:} Beyond driving conventional magnetic undulators, the plasma structure itself can serve as a compact radiation source. This has led to a rich field of research into hybrid and all-plasma light sources:
    
    \begin{itemize}
        \item \textbf{Betatron and Compton Sources:} The strong transverse focusing fields of a plasma bubble compel trapped electrons to undergo betatron oscillations, naturally producing bright, incoherent, broadband X-ray radiation. This process creates a simple, micron-scale source of femtosecond X-rays, ideal for high-resolution imaging \cite{Rousse2004}. Furthermore, by colliding the accelerated electron beam with a counter-propagating laser pulse, one can generate quasi-monoenergetic, tunable gamma-ray beams via inverse Compton scattering, a feat demonstrated in all-optical configurations \cite{TaPhuoc2012NatPhoton}.
        
        \item \textbf{Plasma Undulators:} A more ambitious concept is to use the plasma wakefield itself as a wiggler or undulator to drive an FEL. The idea was pioneered in the 1980s, with foundational work by Yan and Dawson introducing the "ac FEL" concept~\cite{YanDawson1986}, followed by analysis from Joshi \emph{et al.} that established the formal equivalence between plasma wigglers and their magnetic counterparts while also highlighting key performance limitations~\cite{Joshi1987_JQE,Joshi1987_PAC}. Plasma undulators offer the prospect of extremely short periods and large strength parameters ($K$).
        
        While early experiments demonstrated spontaneous X-ray emission from betatron motion in ion channels~\cite{Wang2002_PRL}, achieving coherent FEL gain in a plasma undulator remains a challenge. The primary hurdles include maintaining phase stability over many periods and controlling the radiation bandwidth. Recent proposals are exploring novel ways to overcome these issues, for example, by using plasma channels driven by high-order laser modes with sophisticated phase-locking and tapering techniques~\cite{Rykovanov2016_PRAB, Wang2017_SciRep}. Success in this area would enable fully integrated, all-plasma accelerators and light sources.
    \end{itemize}
\end{itemize}

\subsubsection{Transforming Medicine and Industry}

The potential impact of compact accelerators extends deeply into medical and industrial realms.
\begin{itemize}
    \item \textbf{Advanced Radiotherapy:} Current cancer radiotherapy facilities are large and expensive. Compact plasma accelerators would enable more widespread and advanced forms of treatment. The ultrashort, intense dose delivery of a plasma-accelerated beam is ideally suited for the "FLASH effect", where high doses delivered within a fraction of a second have been shown in vivo to kill tumor cells while sparing healthy tissue more effectively than conventional radiation \cite{Favaudon2014FLASH}. Furthermore, compact proton and ion accelerators driven by plasma-based schemes could make hadron therapy, a more precise form of radiotherapy, much more accessible.

    \item \textbf{Ultrafast Science and Industrial Inspection:} The femtosecond-scale electron bunches and X-ray pulses from plasma accelerators are perfect tools for "filming" molecular and atomic processes in real-time, with applications in chemistry, biology, and materials science. Industrially, compact, high-energy X-ray and neutron sources based on this technology could be used for non-destructive testing of critical components, such as inspecting aircraft engines or nuclear waste containers, with unprecedented resolution and speed.
\end{itemize}

\subsubsection{The Ultimate Goal: A High-Energy Physics Collider}

The ultimate ambition for plasma accelerator technology is to power a future high-energy electron-positron linear collider. By replacing the metallic RF structures with plasma cells, the length of a TeV-scale collider could be reduced from tens of kilometers to perhaps a few hundred meters. This is the primary long-term goal that drives much of the fundamental research in the field. Realizing this vision requires overcoming immense challenges, including staging many plasma cells with high efficiency, preserving beam emittance, accelerating positrons, and achieving extremely high repetition rates and power efficiency. This grand challenge is the focus of a coordinated global effort.

\subsection{The Global Roadmap for PWFA}
\label{sec:roadmap}

Powered by rapid progress over the last decade, the worldwide PWFA effort has matured from single-shot proofs of principle into facility-scale programs with explicit performance milestones and application targets. This global endeavor is now guided by strategic roadmaps that define the long-term vision, underpinned by concrete engineering milestones at flagship facilities aimed at demonstrating collider-relevant parameters.

\subsubsection{Strategic Vision: The Path to a Collider}

The long-term ambition for PWFA is to power a future high-energy collider. This vision is increasingly being formalized in strategic documents from the international particle physics community. The 2023 U.S. P5 Report, for instance, identifies a $\sim$10~TeV parton center-of-mass (pCM) collider as a key long-range goal and explicitly encourages end-to-end design studies for concepts incorporating wakefield technologies~\cite{P5_2023}.

In response, the community has organized targeted design initiatives. The ALEGRO workshop series has systematically outlined the key challenges and work packages for advanced linear colliders, including PWFA-based scenarios~\cite{ALEGRO2024}. Building on this, the "10 TeV pCM Wakefield Collider Design Study" was launched in 2025 to deliver a self-consistent design blueprint with cost scaling, ensuring interoperability with parallel efforts like the HALHF (Hybrid Asymmetric Linear Higgs Factory) concept~\cite{Gessner2025_10TeV, DPF2025_10TeVUpdate, HALHFNJP2023, HALHFStatus2023}. HALHF itself proposes an innovative architecture that combines a PWFA-driven electron beam with an RF-driven positron beam, strategically sidestepping the most difficult positron-in-plasma challenges at the Higgs scale. In Europe, the CERN LDG Accelerator R\&D Roadmap recognizes plasma wakefield acceleration as one of its core pillars, underscoring the need for sustained investment beyond 2026~\cite{EuPRAXIALDGReview2025}.

The UK community published a dedicated, community-driven roadmap in 2019 that sets out a vision for plasma wakefield accelerator research in the years 2019-2040, including specific recommendations: (i) sustained support for university-scale laboratories (e.g. SCAPA); (ii) strong UK participation in EuPRAXIA; and (iii) development of a dedicated PWFA beamline at CLARA/ASTeC. This roadmap complements UK efforts toward PWFA--FEL coupling studies at CLARA and aligns with European initiatives. See Refs.~\cite{Hidding_UKRoadmap2019,PWASC_UKRoadmapWeb,Hanahoe_IPAC14,Xia_NIMA2014,Xia_NIMA2016,CLARA_ASTeC,UKRI_PWFAFEL}.

In parallel, the PWFA landscape in Asia is characterized by a blend of historical contributions, theoretical leadership, and strategic integration with existing national laboratories. As early as 1990, KEK in Japan performed some of the first beam-driven PWFA experiments, demonstrating energy modulation and resonant effects with a 250~MeV linac~\cite{Nakajima1990NIMA, KEK_EPAC1990}. More recently, Asian research groups have become key players in theoretical advancements, particularly on the critical positron acceleration challenge. Notably, a series of papers from 2021--2022 systematically developed the framework for high-efficiency, uniform beam loading of positrons in hollow plasma channels, providing an actionable path toward high-quality positron acceleration~\cite{ZhouPRL2021, Zhou2022PRAB}.

On the experimental front, the focus is on leveraging the capabilities of major existing facilities. In China, community documents highlight a plasma-based injector path toward CEPC~\cite{WeiLu_ALEGRO2019}. At IHEP, designs are underway to adapt the BEPCII linac as a high-quality driver for external injection experiments, with plans to compress its 2~GeV electron bunches for PWFA and LWFA tests~\cite{ShiLINAC2024_BEPCLINE}. 

In Japan, facilities like SPring-8/SACLA offer a mature infrastructure for co-locating advanced accelerator tests with user-grade light sources, demonstrating sophisticated beam switching and synchronization capabilities that are essential for future PWFA-FEL coupling~\cite{MaesakaIPAC2021}. These efforts, often coordinated internationally through frameworks like ALEGRO~\cite{MuggliIPAC2018}, position Asia to play a crucial role in validating key technologies for staging, beam matching, and user facility integration.

\subsubsection{Engineering Milestones: Toward Collider-Relevant Parameters}

Translating the strategic vision into reality requires achieving and integrating several key performance metrics that are critical for a collider. Recent progress at dedicated facilities has provided crucial "reality checks" on these fronts:

\begin{itemize}
    \item \textbf{High-Repetition-Rate Operation:} A collider requires MHz-level repetition rates, a major challenge due to plasma recovery times and thermal loading. Foundational experiments at facilities like FLASHForward and SPARC\_LAB have quantitatively measured sub-nanosecond recovery times in hydrogen plasmas, providing a critical baseline for high-average-power designs~\cite{Darcy2022Recovery, Pompili2024Recovery}.

    \item \textbf{High Energy-Transfer Efficiency:} Efficiently transferring energy from the drive beam to the witness beam is paramount for the overall power efficiency of a collider. Recent experiments have successfully demonstrated high efficiency and have begun to explore the physics of driver beam depletion and even re-acceleration within the plasma, offering pathways for multi-stage energy management and beam loading compensation~\cite{PenaPRResearch2024}.

    \item \textbf{Beam Quality Preservation:} Perhaps the most critical requirement for a collider is preserving the beam's ultra-low emittance throughout the acceleration process. A landmark demonstration at FLASHForward in 2024 achieved full preservation of slice emittance in a PWFA stage, confirming that plasma accelerators are compatible with the stringent beam quality demands of applications like FELs and future colliders~\cite{Lindstrom2024}.
    
    \item \textbf{Staging and Scalability:} Achieving TeV energies will require staging hundreds or thousands of plasma cells. Key milestones include demonstrating efficient staging with emittance preservation (a primary goal for FACET-II) and developing scalable plasma sources capable of operating over hundreds of meters (a central objective for AWAKE Run-2).
\end{itemize}

\subsubsection{Landscape of Facilities: Current Status and Near-Term Goals}

The global PWFA R\&D effort is distributed across several flagship user facilities, each tackling distinct but complementary aspects of the collider challenge. Table~\ref{tab:facilities} summarizes the key parameters and recent or near-term milestones of these major programs. In the United States, SLAC's \textbf{FACET-II} serves as a versatile testbed for high-gradient, high-efficiency acceleration, with a strong focus on positron solutions and staging. In Europe, DESY's \textbf{FLASHForward} leverages a superconducting linac to push the frontiers of beam quality preservation and high-repetition-rate operation. CERN's \textbf{AWAKE} experiment pioneers the use of proton drivers to reach unprecedented energy gains in a single stage, targeting TeV-scale energies. Finally, the \textbf{EuPRAXIA} project, with its implementation at SPARC\_LAB, aims to build the first user-oriented FEL driven by a plasma booster, representing a critical step from fundamental research to applied science. These facility-scale programs, complemented by global design studies, form a synergistic network driving the field toward its 
ultimate goals.

\begin{table*}[t]
\centering
\small
\caption{Major PWFA facilities and their recent (publicly documented) milestones and goals. Timelines reflect information available as of the stated publication dates; detailed schedules are subject to updates from each facility.}
\label{tab:facilities}
\setlength{\tabcolsep}{4pt} 
\begin{tabularx}{\textwidth}{p{3.0cm} p{3.2cm} >{\RaggedRight}X >{\RaggedRight}p{3.2cm}} 
\toprule
\textbf{Facility (Region)} & \textbf{Driver/Mode} & \textbf{Recent Milestones / Technical Focus} & \textbf{Timeline/Reference}\\
\midrule
FACET-II (SLAC, US) & e$^-$ PWFA (10~GeV); H$_2$/Li plasma; programmable beam shaping & \emph{First beam–plasma interaction results published}; Focus on \textbf{efficiency}, \textbf{beam quality}, and \textbf{positron solutions} & 2019 Facility Overview~\cite{Yakimenko2019PRAB}; 2024 First Results~\cite{Storey2024PRAB}\\
\addlinespace[0.3em]
FLASHForward (DESY, EU) & e$^-$ PWFA (FLASH LINAC); co-located w/ FLASH user facility & \textbf{Beam quality preservation}~\cite{Lindstrom2024}; \textbf{Driver depletion \& re-acceleration}~\cite{PenaPRResearch2024}; \textbf{High-repetition-rate} R\&D & 2024--2025 Upgrade~\cite{DESYFLASH2025}\\
\addlinespace[0.3em]
AWAKE Run-2 (CERN, EU) & p-driven PWFA (SPS/LHC driver, SM + acceleration dual plasma) & Goals: \textbf{Stable GV/m gradients}, \textbf{emittance preservation}, \textbf{100-m scale plasma sources} & Program Review~\cite{AWAKE_Run2_2022}; Status Report (2024)~\cite{AWAKEStatus2024}\\
\addlinespace[0.3em]
EuPRAXIA @SPARC\_LAB (INFN, EU) & RF linac + PWFA booster staged to VUV/XUV FEL & Design and prototyping for a \textbf{user-grade FEL}; Installation and commissioning for 2025 onwards & Status Update (2023)~\cite{EuPRAXIAStatus2023}\\
\addlinespace[0.3em]
CLARA/ASTeC (STFC, UK) & e$^-$ PWFA (250~MeV); Co-located w/ VELA/CLF & Development of a dedicated PWFA beamline; Focus on \textbf{PWFA--FEL coupling}, diagnostics, and stability & UK Roadmap~\cite{Hidding_UKRoadmap2019}; Facility Status~\cite{CLARA_ASTeC}\\
\addlinespace[0.3em]
BEPCII LINAC \& SPring-8/SACLA (Asia) & External injection from existing user facilities & BEPCII: Design of \textbf{bunch compression line} for PWFA tests. SPring-8: Demonstrated \textbf{beam switching \& sync} for co-location & BEPCII~\cite{ShiLINAC2024_BEPCLINE}; SPring-8~\cite{MaesakaIPAC2021}\\
\addlinespace[0.3em]
\emph{Design Studies} (Global) & HALHF (e$^-$ PWFA + e$^+$ RF, hybrid Higgs factory); 10~TeV pCM Wakefield Collider & 10~TeV initiative: \textbf{Cross-working-group} parameters and costing. HALHF: \textbf{asymmetric collider concept} & HALHF~\cite{HALHFNJP2023,Foster2025PhysicsOpen}; 10~TeV~\cite{Gessner2025_10TeV}; ALEGRO~\cite{ALEGRO2024}\\
\bottomrule
\end{tabularx}
\end{table*}

\FloatBarrier

\subsection{Outlook and Perspective}

From the initial theoretical sparks of the early 1980s to the global network of advanced experimental facilities operating today, the journey of the plasma wakefield accelerator has been a remarkable testament to the power of a single, compelling idea. In this review, we have traversed this landscape, from the foundational theoretical breakthroughs that first established the PWFA as a viable concept, to the experimental triumphs that have demonstrated tremendous energy gains, and onward to the speculative frontiers where these same wakefields may serve as laboratories for quantum gravity. 

Reflecting on this nearly fifty-year journey, what is perhaps most striking is the persistent, symbiotic evolution between theory and experiment. Foundational concepts like the high transformer ratio and the plasma lens were not merely theoretical curiosities; they became guiding principles that motivated and shaped decades of experimental work. Conversely, the experimental discovery of the pristine, self-organizing nature of the nonlinear bubble regime provided profound new insights that, in turn, spurred a new generation of more sophisticated theoretical models. This virtuous cycle has been the engine of our field's progress.

Looking forward, the path is both challenging and exhilarating. The strategic roadmaps and engineering milestones laid out by the international community provide a clear direction, moving from near-term applications like compact FELs towards the grand challenge of a plasma-based linear collider. The technological hurdles remain formidable, demanding unprecedented control over beam and plasma parameters with ever-increasing precision and repetition rates. Yet, as the relentless progress documented in Table~\ref{tab:facilities} shows, these challenges are being systematically addressed. The spirit of innovation that has characterized this field, giving us every reason to be optimistic.

Perhaps the most profound lesson from this journey is that the plasma wakefield is more than just an accelerator. It is a microcosm of extreme physics, a place where collective phenomena, relativistic dynamics, and even quantum field theory and black hole physics converge in a single, tabletop system. It is a laboratory where we can not only forge the tools to explore the next energy frontier in particle physics but also simulate the most extreme objects such as black holes in the cosmos.

The field is still young, vibrant, and filled with fundamental questions. The grand challenges, from achieving full emittance preservation in a multi-stage collider to making the first detection of analog Hawking radiation, are still waiting to be solved. The journey will undoubtedly require persistence, creativity, and a willingness to work at the confluence of diverse fields. The rewards, both in terms of scientific discovery and technological advancement, will be immense. The dream of a "tabletop TeV" accelerator, once a distant vision, is now a tangible goal on the horizon. The quest continues, and its most exciting chapters are yet to be written.

\bibliography{reference}
\end{document}